\documentclass[11pt,a4paper]{article}

\usepackage{jheppub}
\usepackage{subfigure}
\usepackage{bbm}
\usepackage[T1]{fontenc}
\newcommand{\MSb}{\overline{\textrm{MS}}}
\def\RM{\mathbbm{R}_M}
\def\PM{\mathbbm{P}_M}
\def\dirac#1{\gamma_{#1}}
\def\Tr{{\rm Tr}}

\title{
{\footnotesize\vspace*{-3.5cm}\hspace*{13cm}{\textnormal{DESY 13-208}}\vspace*{-0.4cm}}
{\footnotesize\hspace*{13cm}\textnormal{HU-EP-13/64}\vspace*{-0.4cm}}
{\footnotesize\hspace*{12.6cm}\textnormal{SFB-CPP-13-114}\vspace*{0.5cm}}
Topological susceptibility from the twisted mass Dirac
operator spectrum}

\author[a,b]{Krzysztof Cichy,}
\author[a,c]{Elena Garcia-Ramos,}
\author[a,d]{Karl Jansen}

\affiliation[a]{NIC, DESY, Platanenallee 6, 15738 Zeuthen, Germany}
\affiliation[b]{Adam Mickiewicz University, Faculty of Physics,
Umultowska 85, 61-614 Poznan, Poland}
\affiliation[c]{Humboldt Universit\"at zu Berlin, Newtonstr. 15, 12489
  Berlin, Germany}
\affiliation[d]{Department of Physics, University of Cyprus, P.O. Box 20537, 1678 Nicosia,
Cyprus}

\emailAdd{krzysztof.cichy@desy.de}
\emailAdd{elena.garcia.ramos@desy.de}
\emailAdd{karl.jansen@desy.de}

\abstract{
We present results of our computation of the topological susceptibility with
$N_f=2$ and $N_f=2+1+1$ flavours of maximally twisted mass fermions,
using the method of spectral projectors.
We perform a detailed study of the quark mass dependence and discretization effects. We make an
attempt to confront our data
with chiral perturbation theory and extract the chiral condensate from the quark mass
dependence of the topological susceptibility.  We compare the value with the results of our direct
computation from the slope of the mode number.
We emphasize the role of autocorrelations and the necessity of long Monte Carlo runs to obtain results
with good precision.
We also show our results for the spectral
projector computation of the ratio of renormalization constants $Z_P/Z_S$.
\begin{center}
\vspace*{1cm}
\includegraphics
[width=0.2\textwidth,angle=0]
{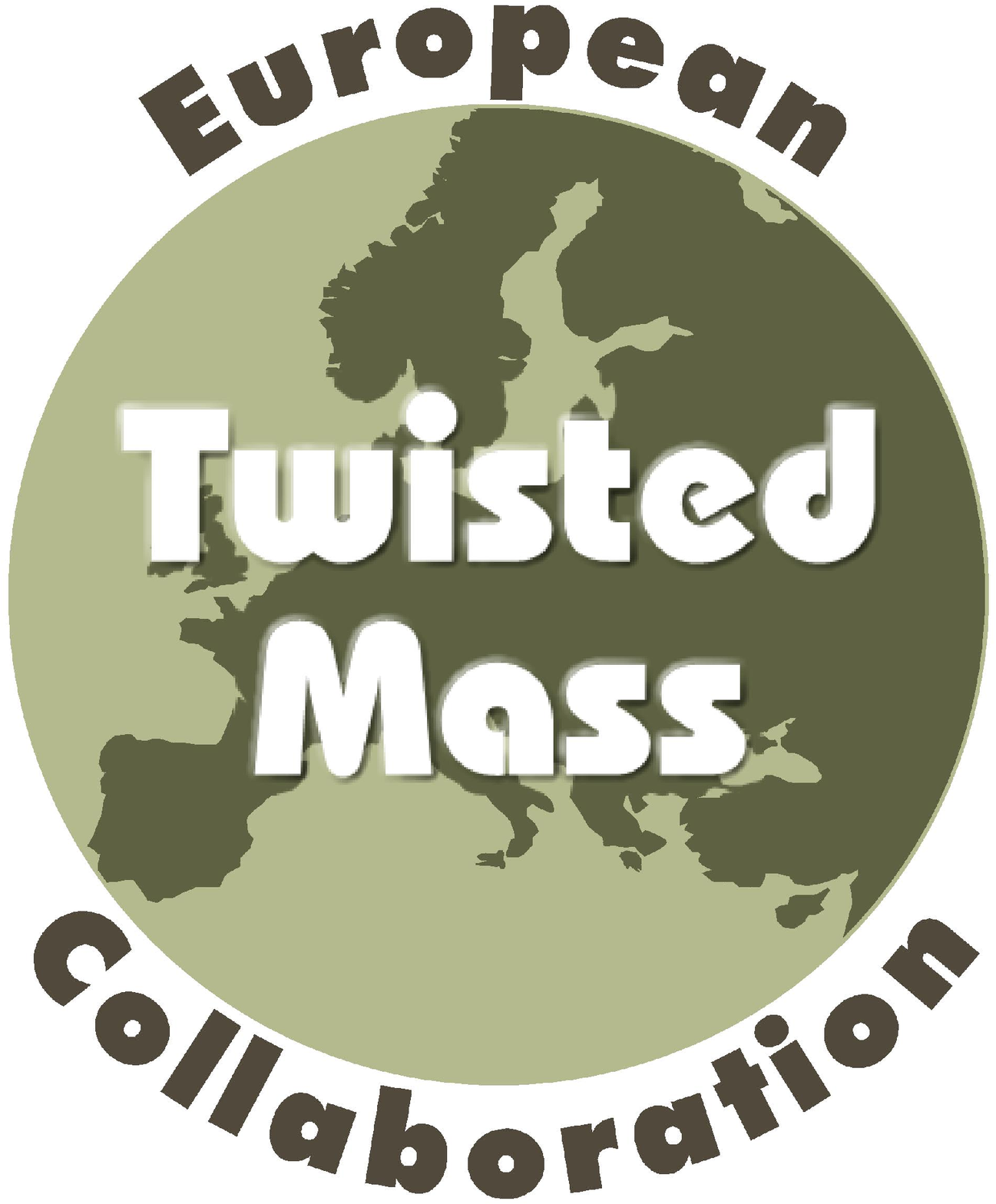}
\end{center}
}

\begin{document}
\maketitle

\section{Introduction}
The topological susceptibility in gauge theories, e.g. in QCD, expresses the fluctuations
of the topological charge.
As such, it describes non-trivial topological properties of the underlying gauge field
configurations.
Such properties have far-reaching phenomenological consequences, in particular topological effects
are to a large extent responsible for the mass of the flavour-singlet pseudoscalar $\eta'$ meson, making
it distinct from the octet of pions, kaons and $\eta$.
The relation between the topological susceptibility and the $\eta'$ mass is expressed in the
Witten-Veneziano formula \cite{Witten:1979vv,Veneziano:1979ec}.

There exist many definitions of the topological charge on the lattice\footnote{For a short review and
discussion of different definitions and for further references, we refer to
Ref.~\cite{Creutz:2010ec}.} and there has been a debate in the literature about the
validity
of different approaches. One of the main problems is the appearance of non-integrable
short distance singularities in some definitions, which require regularization.

To avoid such theoretical problems, a possible solution is to use the definition of the topological
charge as the index of the overlap Dirac operator \cite{Hasenfratz:1998ri}, which is by construction
integer-valued.
However, this is very demanding in terms of computing time and hence impractical when large lattice
sizes are used.
Using Ginsparg-Wilson fermions, it is also possible to derive an expression for the topological
susceptibility which does not have any power divergences \cite{Giusti:2001xh,Giusti:2004qd}.
This has been further generalized by L\"uscher, leading to a definition employing the so-called
density chain correlation functions \cite{Luscher:2004fu}.
The latter can be evaluated efficiently using the method of spectral projectors \cite{Giusti:2008vb}.
This definition of the topological susceptibility was subject to numerical analysis
in the quenched case \cite{Luscher:2010ik}
and it is the aim of the present paper to analyze the results of its usage in the case
with $N_f=2$ and $N_f=2+1+1$ active flavours of twisted mass fermions.

The outline of the paper is as follows. 
In section 2, we describe the theoretical principles of the adopted approach.
Section 3 presents our lattice setup. 
In section 4, we show our results for the renormalization constants ratio $Z_P/Z_S$ and in section 5 for
the topological susceptibility.
We conclude in section 6.
In an appendix, we show our tests concerning the number of stochastic sources.

\section{Theoretical principles}
The method that we follow in this paper was introduced in Refs.~\cite{Giusti:2008vb,Luscher:2010ik}
and we refer to these papers for a comprehensive description. Here, we summarize only the main points to
render the paper self-contained.

Let us define an orthogonal projector $\PM$ to the subspace of fermion
fields spanned by the lowest lying eigenmodes of the operator $D^\dagger D$ with eigenvalues below
some threshold value $M^2$.
In practice, if the projector $\PM$ is approximated by a rational
function of $D^\dagger D$, denoted by $\RM$ (see Refs.~\cite{Giusti:2008vb,Luscher:2010ik} for
the details of this approximation), the following equation for the topological
susceptibility $\chi$ holds:
\begin{equation}
\label{eq:chi}
 \chi={\langle\Tr\{\RM^4\}\rangle\over \langle\Tr\{\dirac{5}\RM^2\dirac{5}\RM^2\}\rangle}
  {\langle\Tr\{\dirac{5}\RM^2\}\Tr\{\dirac{5}\RM^2\}\rangle\over V}.
\end{equation} 
The calculation of the topological susceptibility from this expression requires an evaluation of
three gauge field ensemble averages.
However, if the value of the scheme- and scale-independent ratio $Z_P/Z_S$ is available from
another computation, the above expression can be rewritten as:
\begin{equation}
\label{eq:chi2}
 \chi=\frac{Z_S^2}{Z_P^2}
  \frac{\langle\Tr\{\dirac{5}\RM^2\}\Tr\{\dirac{5}\RM^2\}\rangle}{V},
\end{equation} 
where the numerator can be expressed using two stochastic observables defined in
Ref.~\cite{Luscher:2010ik}:
\begin{equation}
\label{eq:chi3}
\chi=\frac{Z_S^2}{Z_P^2}
  \frac{\langle{\cal C}^2\rangle-\frac{\langle{\cal B}\rangle}{N}}{V}, 
\end{equation} 
where $N$ is the number of randomly generated pseudofermion fields $\eta_i$ added to the
theory\footnote{We use the Z(2) random noise, i.e. $(\eta_i)_r=(\pm1\pm i)/\sqrt{2}$, where $r$ spans
the set of source degrees of freedom (space-time, colour, spin) and all signs $\pm$ are chosen randomly.}
and
\begin{equation}
\label{eq:C}
  {\cal C}={1\over N}\sum_{k=1}^N
  \left(\RM\eta_k,\dirac{5}\RM\eta_k\right),
 \end{equation}
\begin{equation}
\label{eq:B}
 {\cal B}={1\over N}\sum_{k=1}^N
  \left(\RM\dirac{5}\RM\eta_k,\RM\dirac{5}\RM\eta_k\right).
\end{equation}
The term $\langle {\cal B}\rangle/N$ is a correction to the result given by
$\langle{\cal C}^2\rangle$ needed if the number of stochastic sources $N$ is finite and if one
computes:
\begin{equation}
{\cal C}^2\equiv C_{\{\eta_k\}}\cdot C_{\{\eta_l\}}\equiv
{1\over N}\sum_{k=1}^N \left(\RM\eta_k,\dirac{5}\RM\eta_k\right)\;
{1\over N}\sum_{l=1}^N \left(\RM\eta_l,\dirac{5}\RM\eta_l\right)
\end{equation}
using the same stochastic sources for $C_{\{\eta_k\}}$ and $C_{\{\eta_l\}}$.
In chiral symmetry preserving formulations of Lattice QCD (e.g. using overlap fermions), the observable
${\cal C}$ is just the index $Q$ of the Dirac operator, i.e. the difference in the number of zero
modes with
positive and negative chirality, since $(\eta,\gamma_5\eta)=\pm1$ if $\eta$ is a zero mode and 0
otherwise.
Moreover, in such theories $Z_P=Z_S$ and in the limit $N\rightarrow\infty$ Eq.~\eqref{eq:chi3}
becomes just the well-known formula $\chi=\langle Q^2\rangle/V$.
The distribution of $Q$ is expected to be of the Gaussian type (with $\langle Q\rangle=0$) and the
topological susceptibility is then alternatively given by the width of this distribution.
In theories where chiral symmetry is explicitly broken at finite lattice spacing, e.g. for Wilson
fermions,
the observable ${\cal C}$ is in general non-integer and counts the number of zero modes only
approximately (up to cut-off effects).
However, as we will show, ${\cal C}$ is still compatible with a Gaussian-shaped distribution and the
renormalized
$\mathcal{C}_{ren}\equiv \frac{Z_S}{Z_P}\cal{C}$ can be thought of as a proxy for the topological charge.
As it is well known, the topological charge is an observable which is particularly susceptible to
autocorrelations in Monte Carlo (MC) time \cite{Schaefer:2010aa}.
Hence, to obtain reliable estimates of the topological susceptibility, it is essential that MC
histories are long enough, such that all topological sectors are correctly probed.
Since the observable ${\cal C}$ is strongly related to the topological charge, its autocorrelation time
and the quality of its distribution provides a criterion of MC history being ``long enough''.
In particular, we demand the distribution of ${\cal C}$ to be compatible with a Gaussian and
$\langle{\cal C}\rangle$ should be compatible with
zero.

We have mentioned above that the full renormalized topological susceptibility can be obtained from
expression \eqref{eq:chi}. This means that the ratio of renormalization constants $Z_P/Z_S$ can be
calculated with spectral projectors, as first noticed in Ref.~\cite{Giusti:2008vb}. The formula reads:
\begin{equation}
 \frac{Z_P}{Z_S}=\sqrt{\frac{\langle{\cal B}\rangle}{\langle{\cal A}\rangle}},
\end{equation} 
where ${\cal B}$ is given by Eq.~\eqref{eq:B} and ${\cal A}$ is:
\begin{equation}
\label{eq:A}
 {\cal A}={1\over N}\sum_{k=1}^N
  \left(\RM^2\eta_k,\RM^2\eta_k\right),
\end{equation}
i.e. it is the mode number $\nu(M)$ -- the number of eigenmodes of the operator $D^\dagger D$ with
eigenvalues below the threshold value $M^2$.

\section{Lattice setup}
Our computations were performed using dynamical twisted mass configurations generated by the European
Twisted Mass Collaboration (ETMC), with $N_f=2$ \cite{Boucaud:2007uk,Boucaud:2008xu,Baron:2009wt}
or \mbox{$N_f=2+1+1$} \cite{Baron:2010bv,Baron:2010th,Baron:2011sf} dynamical quark flavours.
In the gauge sector, the action is:
\begin{equation}
 S_G[U] = \frac{\beta}{3}\sum_x\Big( b_0 \sum_{\substack{
      \mu,\nu=1\\1\leq\mu<\nu}}^4 \textrm{Re\,Tr} \big( 1 - P^{1\times
1}_{x;\mu,\nu}
\big) 
+ b_1 \sum_{\substack{
      \mu,\nu=1\\\mu \ne \nu}}^4 \textrm{Re\,Tr}\big( 1 - P^{1 \times 2}_{x; \mu, \nu} \big)
\Big),
\end{equation}
with $\beta=6/g_0^2$, $g_0$ the bare coupling and $P^{1\times
1}$, $P^{1\times 2}$ are the plaquette and rectangular Wilson loops, respectively.
For the $N_f=2$ case, the tree-level Symanzik improved action \cite{Weisz:1982zw} was used, i.e. $b_1
= -\frac{1}{12}$ (with the normalization condition $b_0=1-8b_1$), while
in the $N_f=2+1+1$ case, the Iwasaki action \cite{Iwasaki:1985we,Iwasaki:1996sn} was employed, i.e.
$b_1=-0.331$.

The Wilson twisted mass fermion action for the light, 
up and down quarks for both the $N_f=2$ and $N_f=2+1+1$ cases, is
given in the twisted basis by:
\cite{Frezzotti:2000nk,Frezzotti:2003ni,Frezzotti:2004wz,Shindler:2007vp}
\begin{equation}
 S_l[\psi, \bar{\psi}, U] = a^4 \sum_x \bar{\chi}_l(x) \big( D_W + m_0 + i \mu_l \gamma_5 \tau_3
\big)
\chi_l(x),
 \label{tm_light}
\end{equation}
where $\tau^3$ acts in flavour space and $\chi_l=(\chi_u,\,\chi_d)$ is a
two-component vector in flavour space, related to the one in the physical basis by a chiral rotation.
$m_0$ and $\mu_l$ are the bare untwisted and twisted light quark masses, respectively.
The renormalized light quark mass is $\mu_R=Z_P^{-1}\mu_l$.
The standard massless Wilson-Dirac operator $D_W$ reads:
\begin{equation}
 D_W = \frac{1}{2} \big( \gamma_{\mu} (\nabla_{\mu} + \nabla^*_{\mu}) - a \nabla^*_{\mu} \nabla_{\mu}
\big),
\end{equation}
where $\nabla_{\mu}$ and $\nabla^*_{\mu}$ are the forward and backward covariant
derivatives.

The twisted mass action for the heavy doublet is: \cite{Frezzotti:2004wz,Frezzotti:2003xj}
\begin{equation}
 S_h[\psi, \bar{\psi}, U] = a^4 \sum_x \bar{\chi}_h(x) \big( D_W + m_0 + i \mu_\sigma \gamma_5
\tau_1 + \mu_\delta \tau_3
\big)
\chi_h(x),
 \label{tm_heavy}
\end{equation}
where $\mu_\sigma$ is the bare twisted mass with the twist along the $\tau_1$ direction
and $\mu_\delta$ the mass splitting along the $\tau_3$ direction that makes the strange
and charm quark masses non-degenerate.
The physical renormalized strange $m^{s}_R$ and charm $m^{c}_R$ quark masses are related to the bare
parameters $\mu_\sigma$ and $\mu_\delta$
via $m^s_R=Z_P^{-1}\left(\mu_\sigma-(Z_P/Z_S)\mu_\delta\right)$ and
$m^c_R=Z_P^{-1}\left(\mu_\sigma+(Z_P/Z_S)\mu_\delta\right)$.
The heavy quark doublet in the twisted basis $\chi_h=(\chi_c,\,\chi_s)$ is related to the one
in the physical basis by a chiral rotation. 

\begin{table}[t!]
  \centering
  \begin{tabular}[]{ccccccccc}
    Ensemble & $\beta$ & lattice & $a\mu_l$ & $\mu_R$ [MeV] &
    $\kappa_c$  & $L$ [fm]& $m_\pi L$\\
\hline
   b$40.16 $  & 3.90 & $16^3\times32$   & 0.004   & 21 &0.160856   & 1.4 & 2.5 \\
  b$40.20 $  & 3.90 & $20^3\times40$  & 0.004 &  21&  0.160856   &  1.7 & 2.8\\
  b$40.24 $  & 3.90 & $24^3\times48$  & 0.004 &  21&  0.160856   &  2.0 & 3.3\\
  b$40.32$ &  3.90 & $32^3\times64$  & 0.004 & 21& 0.160856 & 2.7 & 4.3 \\
 b$64.24$ & 3.90 & $24^3\times48$  & 0.0064 &34 &0.160856 & 2.0 & 4.1\\
 b$85.24$ & 3.90 & $24^3\times48$  & 0.0085 & 45& 0.160856 & 2.0 & 4.7\\
  c$30.20$ & 4.05 & $20^3\times40$  & 0.003 & 19 &0.157010 &  1.3 & 2.4\\
  d$20.24$ & 4.20 & $24^3\times48$  & 0.002 & 15 &0.154073 &  1.3 & 2.4\\
  e$17.32$ & 4.35 & $32^3\times64$  & 0.00175 & 16 &0.151740 & 1.5 & 2.4\\
  \end{tabular}
  \caption{Parameters of the $N_f=2$ gauge field ensembles
\cite{Boucaud:2007uk,Boucaud:2008xu,Baron:2009wt}. We show the inverse bare coupling $\beta$,
lattice size $(L/a)^3\times(T/a)$, bare twisted light quark mass $a\mu_l$,
renormalized quark mass $\mu_R$ in MeV, critical value of the hopping parameter at which the
PCAC mass vanishes and physical extent of the lattice $L$ in fm and the product $m_\pi L$.}
  \label{setupNf2}
\end{table}

The twisted mass formulation yields an automatic $\mathcal{O}(a)$
improvement of $\mathcal{R}_5$-parity-even quantities
if the twist angle is set to $\pi/2$ (maximal twist).
This is achieved by non-perturbative tuning of the hopping
parameter $\kappa = (8+2 a m_0)^{-1}$
to its critical value, at which the PCAC quark mass vanishes
\cite{Frezzotti:2003ni,Chiarappa:2006ae,Farchioni:2004ma,Farchioni:2004fs,Frezzotti:2005gi,
Jansen:2005kk}.

\begin{table}[t!]
   \centering
  \begin{tabular}[]{ccccccccc}
    Ensemble & $\beta$ &       lattice & $a\mu_l$ & $\mu_{l,R}$ [MeV]&
    $\kappa_c$  & L  [fm] & $m_\pi L$\\
\hline
A30.32 &1.90 & $32^3\times 64$  & 0.0030 & 13 & 0.163272  & 2.8 & 4.0\\
A40.20 &1.90 & $20^3\times 40$  & 0.0040 & 17 & 0.163270  & 1.7 & 3.0\\
A40.24 &1.90 & $24^3\times 48$  & 0.0040 & 17 & 0.163270  & 2.1 & 3.5\\
A40.32 &1.90 & $32^3\times 64$  & 0.0040 & 17 & 0.163270  & 2.8 & 4.5\\
A50.32 &1.90 & $32^3\times 64$  & 0.0050 & 22 & 0.163267  & 2.8 & 5.1\\
A60.24 &1.90 & $24^3\times 48$  & 0.0060 & 26 & 0.163265  & 2.1 & 4.2\\
A80.24 & 1.90 & $24^3\times 48$  & 0.0080 & 35 & 0.163260  & 2.1 & 4.8\\
 B25.32 & 1.95 & $32^3\times64$  & 0.0025 & 13& 0.161240 & 2.5 & 3.4\\
  B35.32 &1.95 & $32^3\times64$  & 0.0035 & 18& 0.161240 &2.5 & 4.0\\
  B55.32 & 1.95 & $32^3\times64$  & 0.0055 &28& 0.161236 &2.5 & 5.0\\
  B75.32 &1.95 & $32^3\times64$  & 0.0075 &38 &0.161232 &2.5 & 5.8\\
  B85.24 & 1.95 & $24^3\times48$  & 0.0085  &45 &0.161231 &1.9 & 4.7\\
  D20.48 &  2.10 & $48^3\times96$  & 0.0020 & 12 &0.156357&2.9 & 3.9\\
  D30.48 & 2.10 & $48^3\times96$  & 0.0030 & 19 &0.156355 &2.9 & 4.7\\
  D45.32 & 2.10 & $32^3\times64$  & 0.0045 & 29 & 0.156315 & 1.9 & 3.9\\
   \end{tabular}
  \caption{Parameters of the $N_f=2+1+1$ gauge field ensembles
\cite{Baron:2010bv,Baron:2010th,Baron:2011sf}. We show the inverse bare coupling $\beta$,
lattice size $(L/a)^3\times(T/a)$, bare twisted light quark mass $\mu_l$,
renormalized quark mass $\mu_{l,R}$ in MeV, critical value of the hopping parameter at which the
PCAC mass vanishes, physical extent of the lattice $L$ in fm and the product $m_\pi L$.}
  \label{setupNf211}
\end{table}

\begin{table}[t!]
  \centering
  \begin{tabular}[]{cccccc}
    $N_f$ & $\beta$ & $a$ [fm] & $Z_P(\MSb,\,2\,{\rm GeV})$ & $Z_P/Z_S$ & $r_0/a$\\
\hline
2 & 3.90 & 0.085 & 0.437(7) & 0.639(3) & 5.35(4)\\
2 & 4.05 & 0.067 & 0.477(6) & 0.682(2) & 6.71(4)\\
2 & 4.20 & 0.054 & 0.501(13) & 0.713(3) & 8.36(6)\\
2 & 4.35 & 0.046 & 0.503(13) & 0.740(3) & 9.81(13)\\
2+1+1 & 1.90 & 0.0863 & 0.529(9) & 0.699(13) & 5.231(38)\\
2+1+1 & 1.95 & 0.0779 & 0.504(5) & 0.697(7) & 5.710(41)\\
2+1+1 & 2.10 & 0.0607 & 0.514(3) & 0.740(5) & 7.538(58)\\
\end{tabular}
  \caption{The approximate values of the lattice spacing $a$
\cite{Blossier:2010cr,Baron:2011sf,Jansen:2011vv},
$r_0/a$
\cite{Blossier:2010cr,Baron:2010bv,Ottnad:2012fv,Jansen:2011vv}, the scheme- and
scale-independent renormalization constants ratio $Z_P/Z_S$ and the renormalization constant $Z_P$ in
the $\MSb$ scheme at
the scale of 2 GeV \cite{Constantinou:2010gr,Alexandrou:2012mt,Cichy:2012is,ETM:2011aa,Palao:priv}, for
different
values of $\beta$ and $N_f=2$ and $N_f=2+1+1$ flavours.}
\label{tab:ZP}
\end{table}

The details of the gauge field ensembles considered for this work are presented in Tab.~\ref{setupNf2}
for $N_f=2$ and Tab.~\ref{setupNf211} for $N_f=2+1+1$.
They include lattice spacings from $a\approx0.045$ fm to $a\approx0.085$ fm and up to
5 quark masses at a given lattice spacing.
The renormalized light quark masses $\mu_R$ are in the range from around 15 to 50 MeV.
The values of the renormalization constant $Z_P$ for different ensembles\footnote{For
$N_f=2+1+1$, the mass-independent renormalization constant $Z_P$ is extracted as a chiral limit of a
dedicated computation with 4 mass-degenerate flavours -- see Refs.~\cite{Dimopoulos:2011wz,ETM:2011aa}
for details.}
\cite{Constantinou:2010gr,Alexandrou:2012mt,Palao:priv}, used to convert
bare light quark masses $\mu_l$
and bare spectral threshold parameters $M$ to their renormalized values in the $\MSb$ scheme (at the
scale of 2 GeV), are given in Tab.~\ref{tab:ZP}. There we also show the values of $r_0/a$ (in the
chiral limit), used to
express our results for the topological susceptibility as a dimensionless product $r_0^4\chi$.
Our physical lattice extents $L$ for extracting physical results range from 2 fm to 3 fm (in the
temporal direction, we always have $T=2L$).
To check for the size of finite volume effects, we included different lattice sizes for $\beta=3.9$,
$a\mu_l=0.004$ ($N_f=2$) and $\beta=1.9$, $a\mu_l=0.004$ ($N_f=2+1+1$).

\section{Results -- $Z_P/Z_S$}
\label{sec:ZPZS}
We first present our results for the renormalization constants ratio $Z_P/Z_S$, which is a scale- and
scheme-independent quantity.
Nevertheless, in order to avoid problems with e.g. cut-off effects or dependence on the threshold
parameter $M_R$, it is necessary to determine a window $\Lambda\ll M_R \ll a^{-1}$ for the computation
of $Z_P/Z_S$, with $\Lambda$ of $\mathcal{O}(\Lambda_{\rm QCD})$.

\begin{figure}[t!]
  \begin{center}
    \includegraphics[width=0.9\textwidth,angle=0]{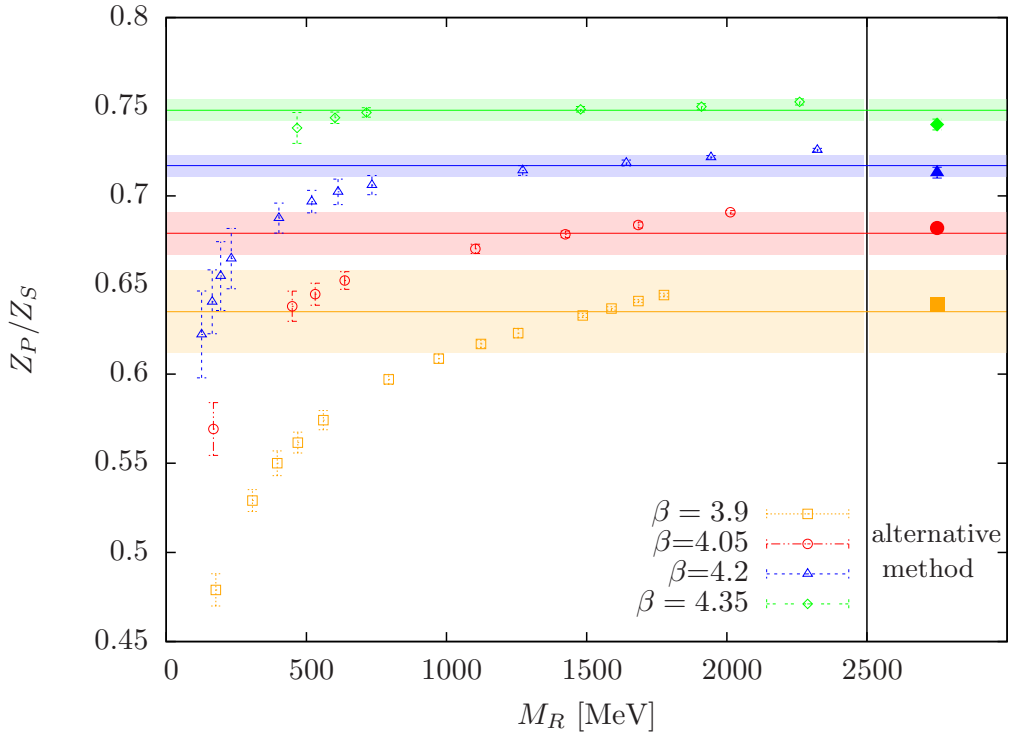}
  \end{center}
  \caption{\label{fig:ZPZS-Nf2} Dependence of the renormalization constants ratio $Z_P/Z_S$ on the
renormalized threshold $M_R$. The data points correspond to the computation from spectral projectors.
The horizontal bands are our estimates of the scale-independent values of $Z_P/Z_S$ that correspond to
the value at $M_R=1.5$ GeV (solid lines) and the spread of results between $M_R=1$ GeV and 2 GeV as our
estimate of the systematic error (bands). The values on the right of the vertical line correspond to
RI-MOM
results at $\beta=3.9,\,4.05,\,4.2$ \cite{Alexandrou:2012mt} and the X-space result at $\beta=4.35$
\cite{Cichy:2012is}.}
\end{figure}

\begin{figure}[t!]
  \begin{center}
    \includegraphics[width=0.9\textwidth,angle=0]{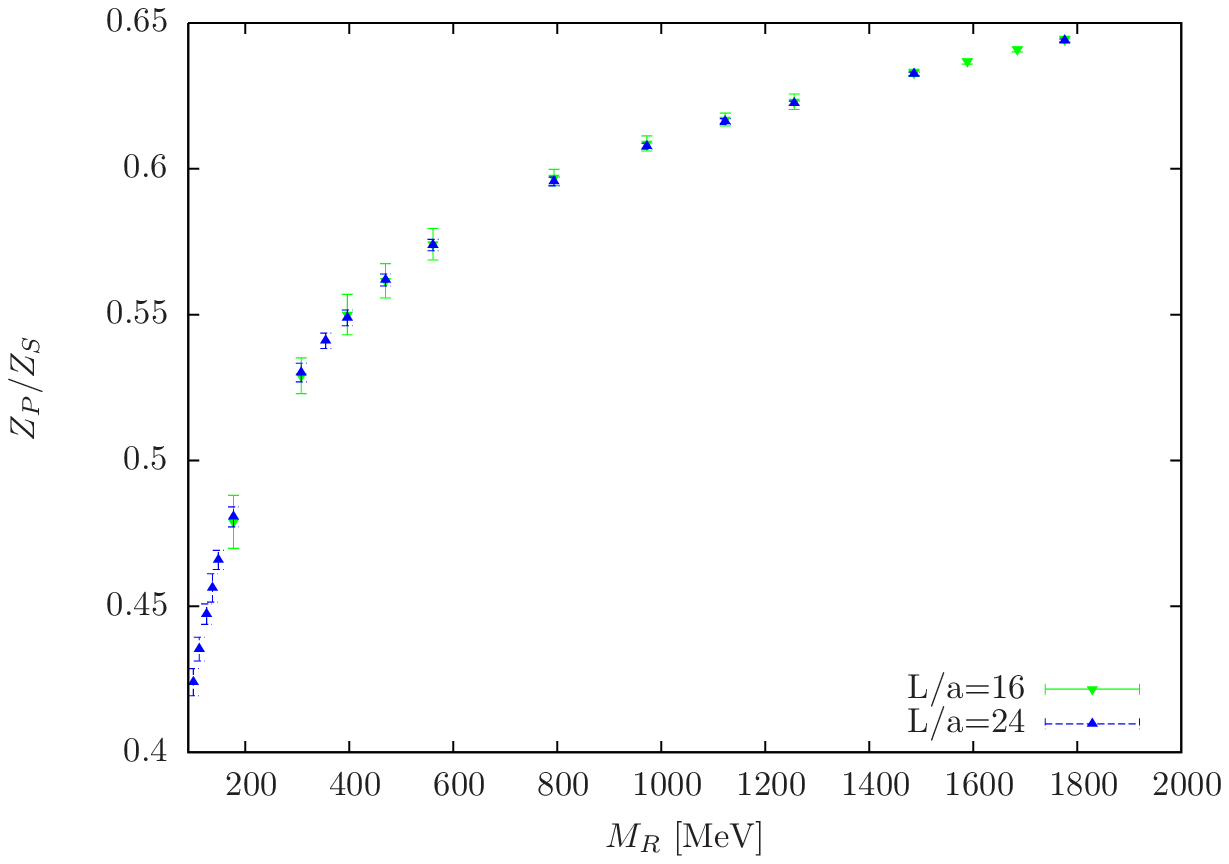}
  \end{center}
  \caption{\label{fig:ZPZS-fve} Dependence of the renormalization constants ratio $Z_P/Z_S$ on the
renormalized threshold $M_R$ for $N_f=2$, $\beta=3.9$, $a\mu_l=0.004$ and two linear lattice extents:
$L/a=16$ and $L/a=24$. Within errors, all values of $Z_P/Z_S$ are compatible between the two ensembles.}
\end{figure}

\subsection{$N_f=2$}
We perform our $N_f=2$ analysis using small volume ensembles (b40.16, c30.20, d20.24 and e17.32)
at a fixed pion mass of around 300 MeV (in infinite volume). In
this way, we can keep the computational cost rather low and at the same time investigate a wide range of
values of $M_R$ to control the systematic effects of varying $M_R$.
For all these ensembles, we can compare the values of $Z_P/Z_S$ with an alternative computation
-- in the
framework of the RI-MOM renormalization scheme ($\beta=3.9,\,4.05,\,4.2$ \cite{Alexandrou:2012mt}) or
the X-space renormalization scheme ($\beta=4.35$ \cite{Cichy:2012is}).

The dependence of $Z_P/Z_S$ on $M_R$ is shown in Fig.~\ref{fig:ZPZS-Nf2}.
For small values of $M_R$, we observe a significant dependence of $Z_P/Z_S$ on the threshold
parameter $M_R$.
For larger values of $M_R$, the dependence of $Z_P/Z_S$ on $M_R$ flattens and we observe a tendency
to approach a plateau.
This signals that we obtain the above discussed lattice window, where we can extract the scale- and
scheme-independent value.
However, the lattice data shows that the strong dependence of $Z_P/Z_S$ on $M_R$ below about 1 GeV is
related to the value of the lattice spacing $a$ involved.
In addition, also the size of variation of $Z_P/Z_S$ for $1\,\textrm{GeV}<M_R<2\,\textrm{GeV}$ is getting
smaller for decreasing values of the lattice spacing.
For $\beta=3.9$, the change in $Z_P/Z_S$ when going from around 1 to 2 GeV is approx. 6\%,
while for finer lattice spacings this change decreases to approx. 3\%, 2\% and below 1\% (for
$\beta=4.05$, $\beta=4.2$ and $\beta=4.35$, respectively).
A reassuring observation is, however, that the plateau value is consistent with values
from the RI-MOM or X-space renormalization schemes, shown in Fig. 1 too.
We remark also that at large values of $M_R$, finite volume effects are small.
We have explicitly checked that the gauge ensemble averages of the observables ${\cal A}$ and
${\cal B}$ are always compatible with each other for ensembles b40.16 and b40.24, provided that
$M_R\gtrsim1$ GeV. Moreover, the ratio of these two quantities, which gives $Z_P/Z_S$, is
compatible between the two ensembles at all values of $M_R$ that we investigated -- see
Fig.~\ref{fig:ZPZS-fve}.
Therefore, the conclusions from our small volume results are valid in general.

To summarize, the method of spectral projectors allows in principle a computation of the ratio of
$Z_P/Z_S$.
However, to obtain the universal scale- and scheme-independent value, the calculation of the observables
${\cal
A}$ and ${\cal B}$ (with $N=1$ stochastic source) has to be performed at a rather large number of
threshold parameters $M_R$ to be able to explore the significant $M_R$-dependence we observe.
To account for this $M_R$-dependence, we followed the strategy to take the central value
of $Z_P/Z_S$ at some fixed (in physical units) value of $M_R$, e.g. 1.5 GeV (sufficiently far away from
the low-energy scales
and sufficiently below the inverse lattice spacing for typical parameters of contemporary simulations),
and assign a systematic error related to the difference of $Z_P/Z_S$ across a range of scales (e.g.
$M_R$ between 1 and 2 GeV). 
If we follow this strategy, we obtain the results of Tab.~\ref{tab:ZPZS} and the horizontal bands in
Fig.~\ref{fig:ZPZS-Nf2}. The first given error of the
spectral projectors result is statistical and the second one comes from the residual $M_R$-dependence of
$Z_P/Z_S$.
Note that only the RI-MOM results, given for comparison in Tab.~\ref{tab:ZPZS}, were chirally
extrapolated.
However, non-zero quark mass corrections to the chiral limit value were found to be small in our setup,
both in the RI-MOM scheme and in the X-space
scheme \cite{Constantinou:2010gr,Alexandrou:2012mt,Cichy:2012is}. In particular, the values of
$Z_P/Z_S$ in the chiral limit and at the pion mass of 300 MeV are always compatible in both of these
schemes. 
Given our experience that $Z_P/Z_S$ has a very mild quark mass dependence
\cite{Constantinou:2010gr,Alexandrou:2012mt,Cichy:2012is}, we consider the values obtained here at a
fixed but small pion mass to be an appropriate estimate.
The overall agreement between the spectral projector method and other renormalization schemes is
certainly reassuring.
However, it would be very good to understand the $M_R$-dependence of $Z_P/Z_S$ better and to disentangle
effects that lead to it. For example, a lattice perturbative calculation within the framework used would
be very helpful to learn about the role of cut-off effects.

\begin{table}[t!]
  \centering
  \begin{tabular}[]{cccc}
    $\beta$ & $Z_P/Z_S$ (spec.proj.) & $Z_P/Z_S$ (RI-MOM)& $Z_P/Z_S$ (X-space)\\
\hline
3.9 & 0.635(1)(23) & 0.639(3) & 0.609(6)\\
4.05 & 0.679(2)(12) & 0.682(2)& 0.671(9)\\
4.2 & 0.717(2)(5) & 0.713(3) & 0.707(14)\\
4.35 & 0.749(2)(2) & -- & 0.740(3)\\  
\end{tabular}
  \caption{The values of the scale- and scheme-independent ratio $Z_P/Z_S$ for $N_f=2$
ensembles, extracted from spectral
projectors (the first error given is statistical and the second one systematic from varying the
threshold value $M_R$), as compared to RI-MOM \cite{Alexandrou:2012mt} and X-space results
\cite{Cichy:2012is}. All RI-MOM results and the X-space results at $\beta=3.9$ and $\beta=4.05$ were
chirally extrapolated.}
\label{tab:ZPZS}
\end{table}

\subsection{$N_f=2+1+1$}
We repeated the computation of the $M_R$-dependence of $Z_P/Z_S$ also for one chosen ensemble with
$N_f=2+1+1$ flavours (B55.32).
The chiral limit value from RI-MOM is 0.697(7) \cite{Palao:priv}.
The residual $M_R$-dependence originating from spectral projectors is rather large in this case and the
prescription
from the previous subsection leads to the value 0.637(1)(21).
The systematic error is comparable to the one for $\beta=3.9$ with $N_f=2$, which corresponds to a
similar lattice spacing.
Although there is some tension between the spectral projector result and RI-MOM, the observed difference
between the two results is still plausible, given a finite and rather large value of the lattice
spacing.
Note, however, that in the following we do not rely on the values of $Z_P/Z_S$ from spectral projectors
-- we rather use
the RI-MOM values to evaluate the topological susceptibility.

\section{Results -- topological susceptibility}
\label{sec:results}
In this section, we discuss our results for the topological susceptibility. We first show the details
of our analysis for two of the 2+1+1-flavour ensembles -- B55.32 and B75.32. Then, we investigate
finite volume effects and finally we present results for the cases with $N_f=2$ and $N_f=2+1+1$ flavours
of twisted mass fermions and perform chiral perturbation theory fits to the quark mass dependence of the
topological susceptibility.

\begin{figure}[t!]
  \begin{center}
   \includegraphics[width=0.345\textwidth,angle=270]{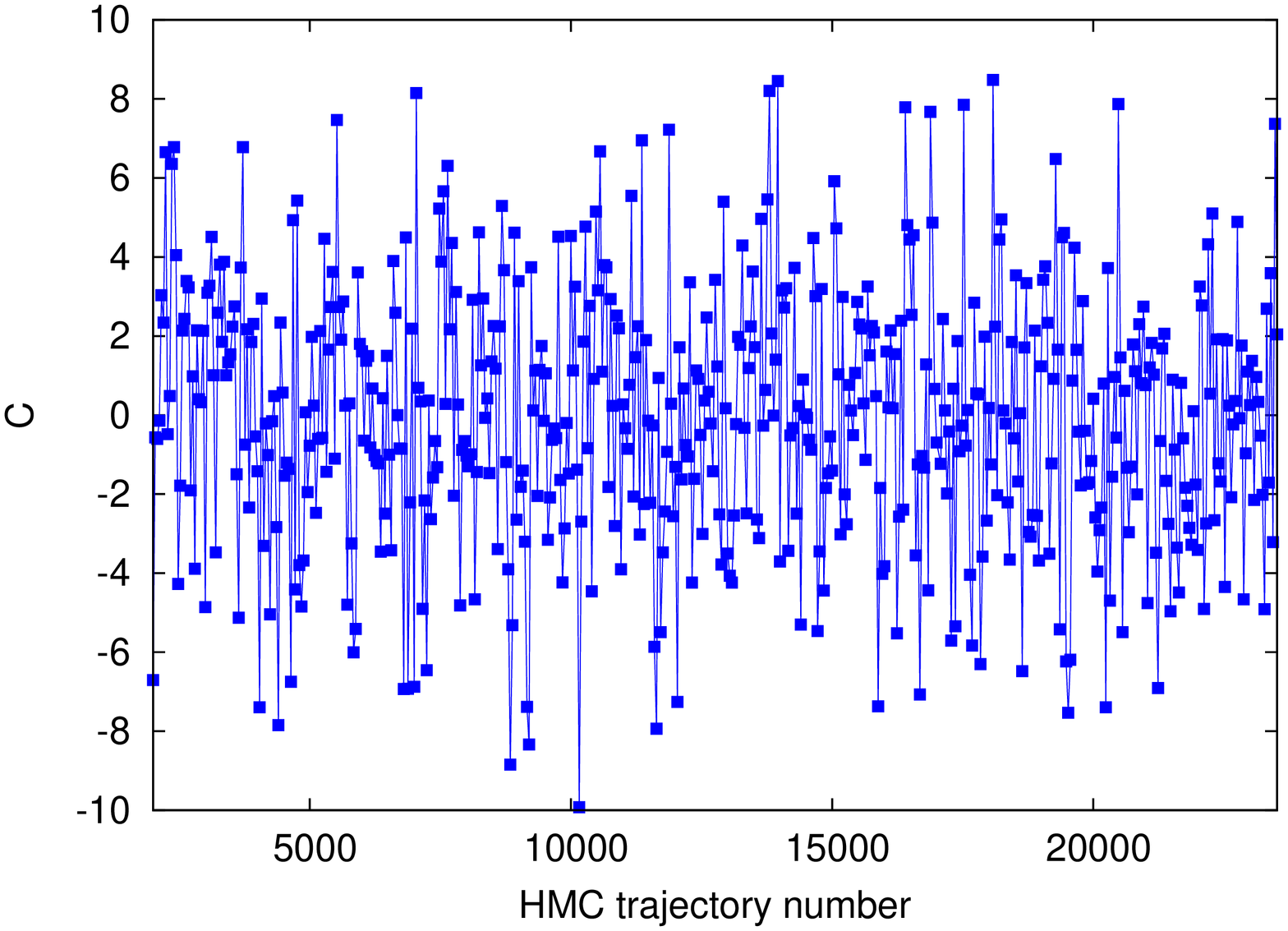}
   \includegraphics[width=0.345\textwidth,angle=270]{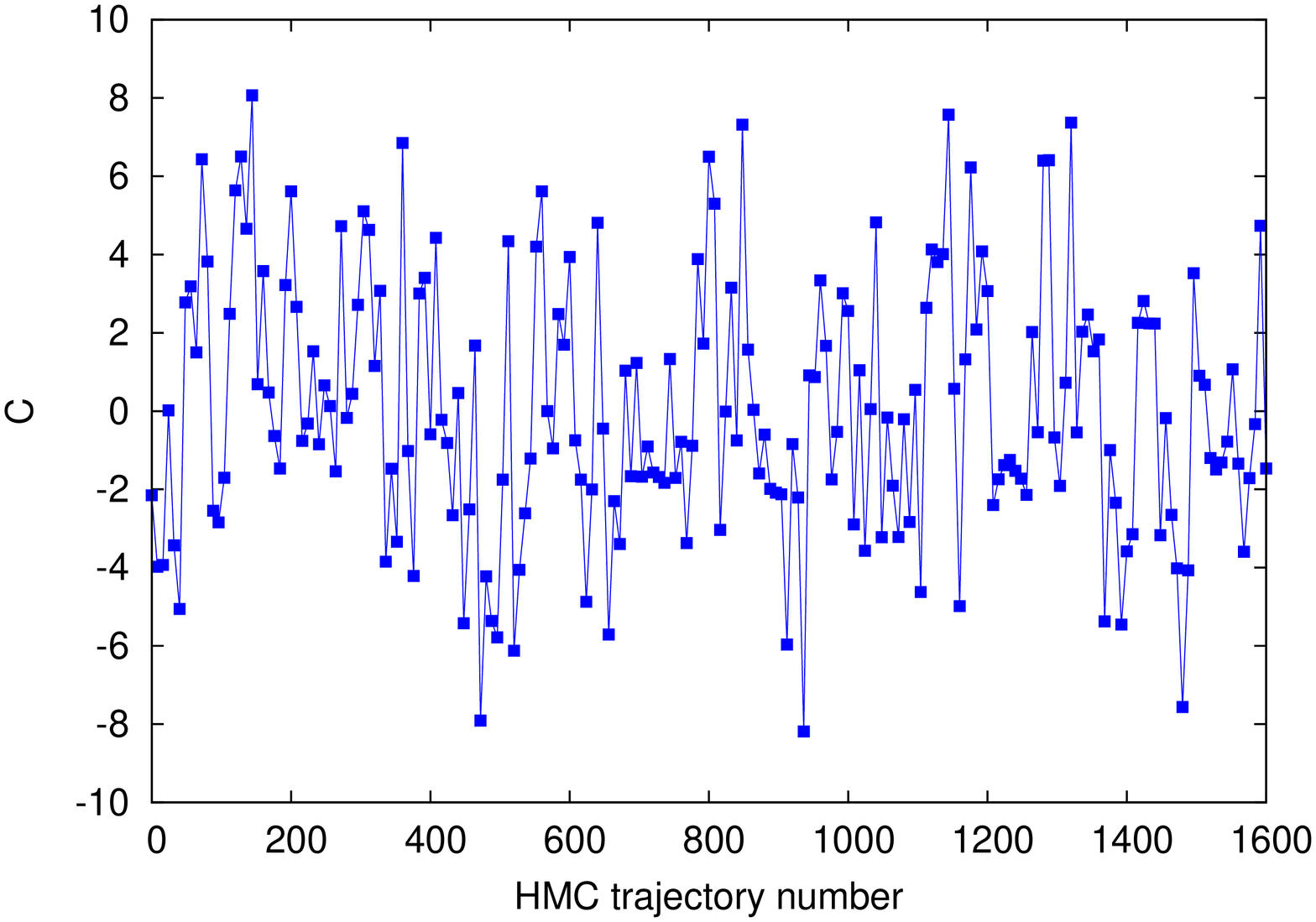}
  \end{center}
  \caption{\label{fig:B55-C} Monte Carlo history of the observable ${\cal C}$ for ensemble B55.32
(left) and B75.32 (right).} 
\end{figure}

\begin{figure}[t!]
  \begin{center}
    \includegraphics[width=0.345\textwidth,angle=270]{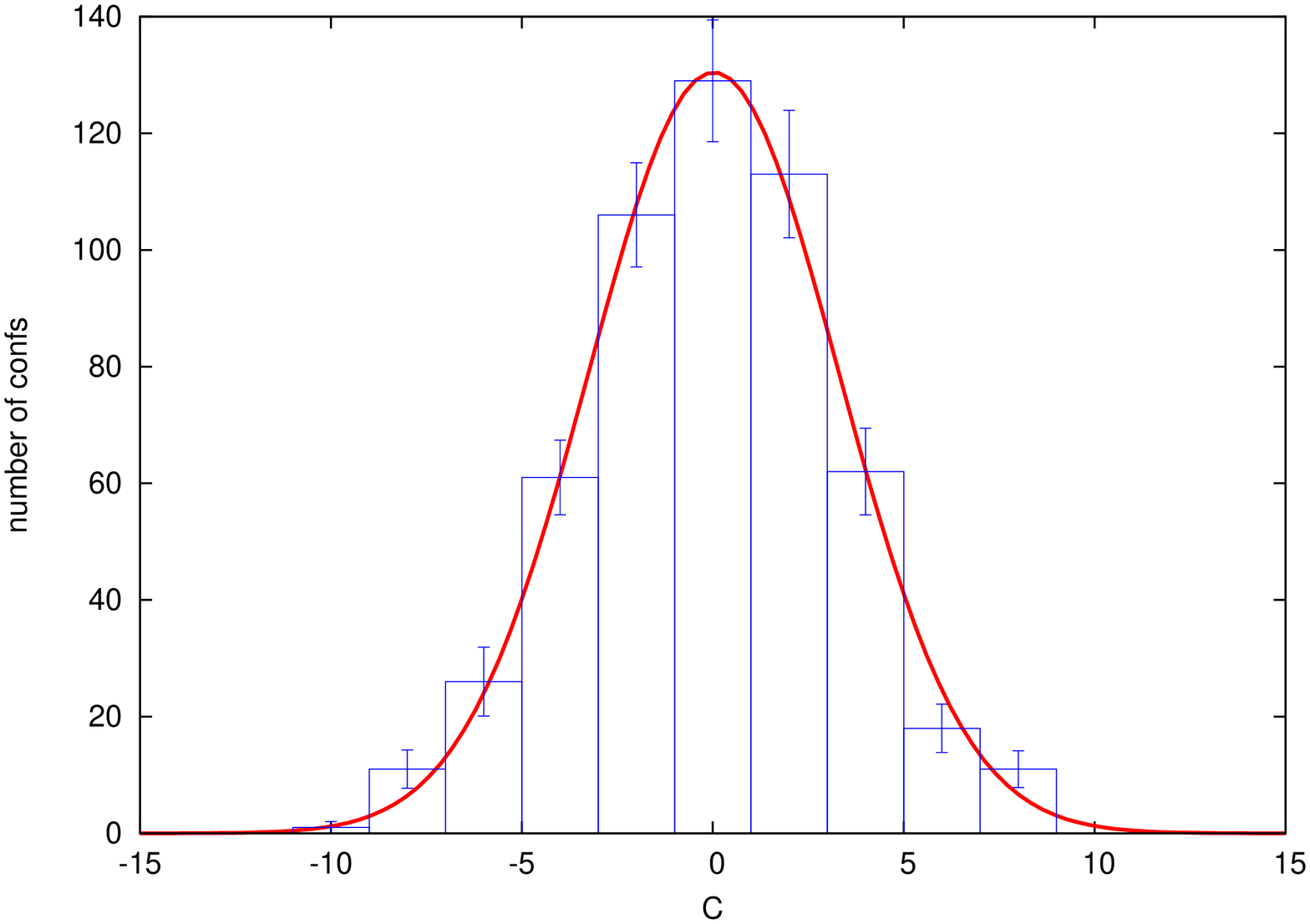}
    \includegraphics[width=0.345\textwidth,angle=270]{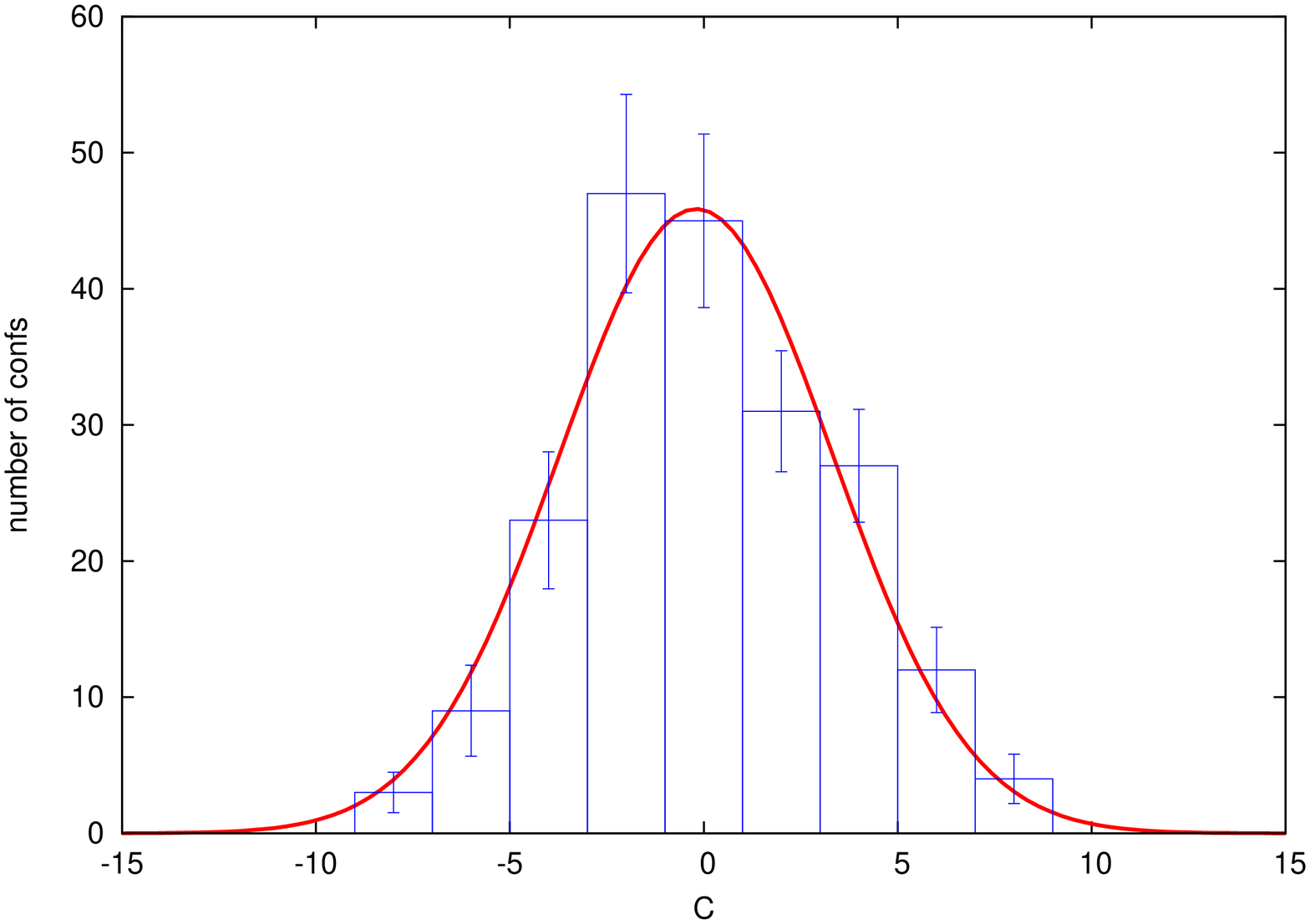}
  \end{center}
  \caption{\label{fig:B55-h} Histogram of the observable ${\cal C}$ for ensemble B55.32 (left) and
B75.32 (right). The error for each box comes from a bootstrap analysis with blocking. The solid line is a
Gaussian fit to the histogram.} 
\end{figure}

\subsection{Examples -- ensemble B55.32 and B75.32}
We start with the ensemble B55.32, see Tab.~\ref{tab:all211}, for which we performed
measurements on 538 independent gauge field configurations separated by 40 MC trajectories, using $N=6$
stochastic sources for each configuration.
For a discussion about the optimal number of stochastic sources per configuration, we refer to Appendix
\ref{sec:Nsrc}.

The MC history of the observable ${\cal C}$ (whose fluctuations determine the topological
susceptibility) is shown in the left panel of Fig.~\ref{fig:B55-C}.
We observe that different topological sectors are sampled and the magnitude of fluctuations seems to be
rather uniform for different regions of MC time.
As we have stated above, the sampling is correct if the histogram of ${\cal C}$ is close to Gaussian and
if the ensemble average $\langle{\cal C}\rangle=0$.
The histogram of the observable ${\cal C}$ for the ensemble B55.32 is shown in Fig.~\ref{fig:B55-h}
(left). It is almost ideally symmetric and it is almost
perfectly Gaussian.
We have therefore fitted the following Gaussian ansatz:
\begin{equation}
 f({\cal C})={\cal N}\exp(-({\cal C}-\langle{\cal C}\rangle)^2/2\sigma^2),
\end{equation}  
where ${\cal N}$ is a normalization constant and $\sigma$ is related to the topological susceptibility:
$\chi=(Z_S/Z_P)^2(\sigma^2-\langle{\cal B}\rangle/N)$, i.e. $\sigma^2=\langle{\cal C}^2\rangle$.
The 3 fitting parameters are then: ${\cal N}$, $\langle{\cal C}\rangle$ and $\sigma$.
There is very good agreement between $\langle{\cal C}\rangle$ extracted from the histogram and
computed directly
by averaging -- the former yields 0.02(20) and the latter -0.06(16), which implies that both the
negative and positive topological charge sectors are sampled equally often.
The bare topological susceptibility extracted from the direct computation and using $\sigma^2$ from the
fit of the histogram is $3.56(51)\cdot10^{-6}$ (histogram) and $3.65(33)\cdot10^{-6}$ (direct).
This agreement implies that indeed the observable ${\cal C}$ is Gaussian distributed and can be
interpreted to play the role of the topological charge.
We also note that the constructed histograms and the extracted values of $\langle{\cal C}\rangle$
and $a^4\chi$ depend very little on the chosen bin size. Using bin sizes of 0.5, 1, 2 and 3, our results
for $a^4\chi$ change only by a few percent and are fully compatible within errors. 
Hence, we decided to use such bin size that the number of bins with non-zero number of gauge field
configurations is around 10.
We emphasize that the good properties of the histogram ($\langle{\cal C}\rangle\approx0$ and Gaussian
shape) hold only if the MC history is long enough. We think that both properties can provide a good
benchmark whether the topological charge sectors are sampled in a correct way.

The statistics that we have for ensemble B55.32 is significantly higher than for other ensembles.
Let us show the details for a more typical ensemble B75.32 with around 100 independent measurements.
The Monte Carlo history (right panel of Fig.~\ref{fig:B55-C}) indicates a correct sampling of
topological sectors, however it is not
long enough to build a fully symmetric histogram (Fig.~\ref{fig:B55-h} (right)).
For example, the number of configurations for which $-3\leq{\cal C}<-1$ and $1\leq{\cal C}<3$ is,
respectively, 47(7) and 31(5), where the error comes from bootstrap with blocking analysis and takes
into account autocorrelations.
Hence, in the generated ensemble, the samples with slightly negative topological charge are somewhat
overrepresented with respect to the ones with slightly positive topological charge, although
statistically they are still compatible.
As a consequence, the peak of the Gaussian fit is for ${\cal C}$ below zero.
Nevertheless, within the computed errors we observe that the shape of
the histogram is close to Gaussian and the topological susceptibility and $\langle{\cal C}\rangle$ are
within large errors compatible between the fit and the direct computation and read for ensemble
B75.32: $\langle{\cal C}\rangle=0.04(35)$ (direct) and -0.20(37) (histogram), bare topological
susceptibility: $a^4\chi=4.13(48)\cdot10^{-6}$ (direct), $a^4\chi=4.80(1.10)\cdot10^{-6}$ (histogram).
However, we would like to give a warning that the rather low statistics we have for the ensemble B75.32
may lead to an underestimation of the error, i.e. the error of the error might be large.
To reach full confidence for the obtained results, statistics of at least the size we have for the
ensemble B55.32 would be necessary.

\subsection{Finite volume effects}
\begin{figure}[t!]
  \begin{center}
    \includegraphics[width=0.65\textwidth,angle=270]{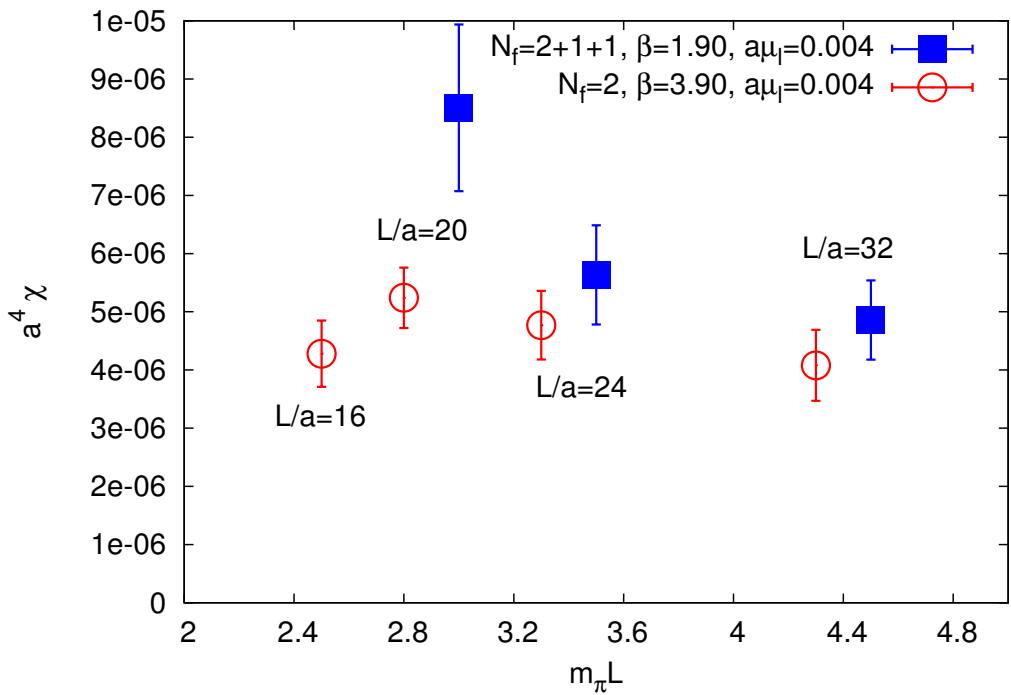}
  \end{center}
  \caption{\label{fig:fve} Bare topological susceptibility for ensembles A40.20,
A40.24 and
A40.32 ($N_f=2+1+1$) and b40.16, b40.20, b40.24 and b40.32 ($N_f=2$).} 
\end{figure}

Before we show results for all our ensembles, we shortly discuss finite volume effects (FVE) in our
simulations.
We show the bare topological susceptibility for three $N_f=2+1+1$ ensembles at $\beta=1.90$,
$a\mu_l=0.004$
and four $N_f=2$ ensembles at $\beta=3.90$, $a\mu_l=0.004$ in Fig.~\ref{fig:fve}. 
All ensembles give compatible results (with some tension between A40.20 and A40.32),
but given the precision we have for these ensembles, i.e. statistical errors of the order of 10-20\%, we
can not conclude about the size of FVE from numerical data.
However, general arguments involving the size of FVE imply that they should be exponentially small if
$m_\pi L\gtrsim4$ (see e.g. Ref.~\cite{Colangelo:2005gd}). 
Since this condition is satisfied for almost all of our $N_f=2+1+1$ ensembles (see the last column of
Tab.~\ref{setupNf211}), we are confident that FVE are much smaller than our statistical errors.
For $N_f=2$, we analyze the quark mass dependence of the topological susceptibility only at one lattice
spacing ($\beta=3.9$) and the product $m_\pi L>4$ for all of them.
However, even in a small volume ($L\approx1.3$ fm, with $m_\pi
L\approx2.4$), FVE are not larger than the statistical errors (cf.
$\beta=3.9$, $L/a=16$ and $L/a=32$ in Fig.~\ref{fig:fve}).

\begin{figure}[t!]
  \begin{center}
    \includegraphics[width=0.9\textwidth,angle=0]{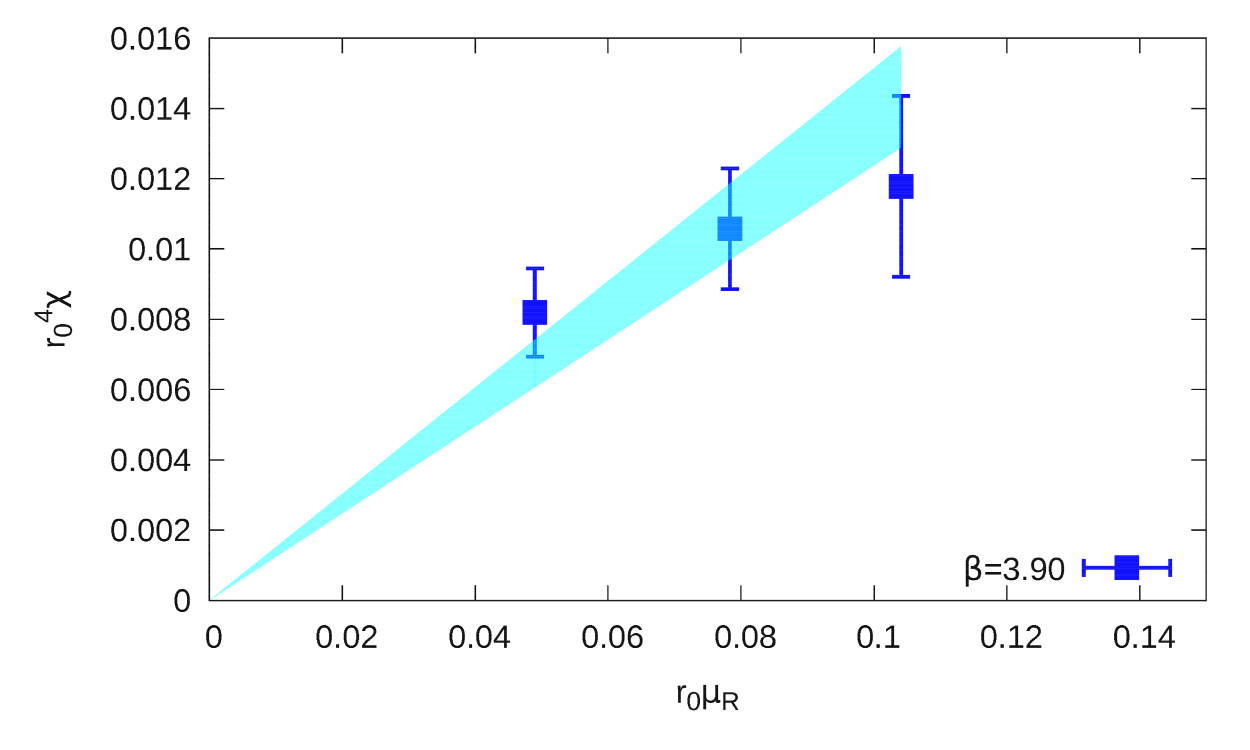}
  \end{center}
  \caption{\label{fig:b39} Renormalized quark mass dependence of the renormalized topological
susceptibility (normalized with $r_0^4$) for $N_f=2$ ensembles at $\beta=3.9$. The fit is to a LO
$\chi$PT expression, $\mathbf{\chi^2/{\rm d.o.f.}\approx1.16}$.} 
\end{figure}

\subsection{$N_f=2$ results}
Tab.~\ref{tab:all2} provides our results for the observables $\langle{\cal A}\rangle$, $\langle{\cal
B}\rangle$, $\langle{\cal C}\rangle$ and the topological susceptibility in the case of $N_f=2$
flavours.
In Fig.~\ref{fig:b39}, we show our results at a single lattice spacing corresponding to
$\beta=3.9$ and a physical volume such that the condition $m_\pi L>4$ is satisfied.
In order to test whether the obtained values of the topological susceptibility could, in principle, be
used to obtain a value for the chiral condensate, we apply the leading order (LO) Chiral Perturbation
Theory ($\chi$PT) expression for $N_f$ flavours of light quarks:
\begin{equation}
\label{xpt}
\chi=\frac{\Sigma\mu_l}{N_f},
\end{equation} 
where $\Sigma$ is the chiral condensate.
In particular, we \emph{impose} that the topological susceptibility vanishes at zero quark mass.
Working with the assumption that LO$\chi$PT can be applied,
the slope of this fit gives the following result for the renormalized condensate ($\MSb$ scheme at 2
GeV):
\begin{displaymath}
 r_0\Sigma^{1/3}=0.650(22),
\end{displaymath}
where the error is mostly statistical, but takes into account also the uncertainties of $Z_P/Z_S$, $Z_P$
and $r_0/a$. The error decomposition is as follows: $r_0\Sigma^{1/3}=0.650(21)(6)(3)(5)$, where
the first error is statistical, the second comes from the uncertainty of $Z_P/Z_S$ (entering via
Eq.~\eqref{eq:chi3}), the third one from the uncertainty of $Z_P$ (entering the renormalized quark mass
$\mu_l$) and the fourth one from the final conversion of $a\Sigma^{1/3}$ to $r_0\Sigma^{1/3}$. The used
values of $Z_P/Z_S$, $Z_P$ and $r_0/a$, together with their uncertainties, are shown in
Tab.~\ref{tab:ZP}. The final error quoted is the sum of the individual errors, combined in quadrature.
We recall here respective values from our direct determination from the mode number of the Dirac
operator: 0.696(20) (at $\beta=3.9$ in the chiral limit) or 0.689(33) (in the continuum limit and in the
chiral limit) \cite{Cichy:2013gja}.
The fact that we observe an agreement indicates a posteriori the validity of our assumption about the
applicability of LO$\chi$PT, at least within the large errors of our present results\footnote{We mention
that we attempted NLO$\chi$PT fits, but the resulting errors on the fit parameters were too large to say
whether higher order corrections are statistically significant.}.

\begin{table}[t!]
\begin{small}
   \centering
  \begin{tabular}[]{ccccccccccc}
    Ens. & $N$ & cnfs & step & $\langle A\rangle$ & $\tau_{int}$ & $\langle B\rangle$
& $\tau_{int}$ & $\langle C\rangle$ & $\tau_{int}$ & $r_0^4\chi$\\
\hline
b40.16 & 12 & 272 & 20 & 5.29(14) & 1.9(6) & 0.92(4) & 2.0(6) & -0.19(9) & 1.6(5) & 0.0097(16)(1)(3)\\ 
b40.20 & 6 & 264 & 20 & 14.61(38) & 3.5(1.3)& 2.59(8) & 3.1(1.1) & -0.10(12) & 0.9(2)  &
0.0092(11)(1)(3)\\
b40.24 & 6 & 454 & 20 & 32.07(19) & 1.0(2) & 5.72(5) & 1.1(2) & -0.13(17) & 1.5(4)  & 0.0096(11)(1)(3)\\
b40.32 & 12 & 217 & 16 & 100.5(5) & 1.7(6) & 17.76(11) & 1.4(4) & -0.38(37) & 1.5(5) &0.0082(13)(1)(3)\\
b64.24 & 6 & 219 & 20 & 30.94(28) & 1.0(3) & 5.39(7) & 1.2(4) & -0.02(27) & 1.8(6) & 0.0106(17)(1)(3)\\
b85.24 & 6 & 160 & 20 & 29.30(24) & 0.6(1) & 5.03(6) & 0.8(2) & 0.47(29) & 1.4(5) & 0.0118(25)(1)(3)\\
\hline
   \end{tabular}
  \caption{Our results for $N_f=2$ flavours. We give the ensemble label, the number of stochastic
sources $N$, the number of configurations used (cnfs), the step between measurements (in units of
molecular dynamics trajectories) and the values of
$\langle{\cal A}\rangle$, $\langle{\cal B}\rangle$, $\langle{\cal C}\rangle$ and the topological
susceptibility, together with integrated autocorrelation times $\tau_{int}$ (in units of
measurements).
The error for $r_0^4\chi$ is, respectively, statistical,
resulting from the uncertainty of $Z_P/Z_S$ (from the RI-MOM method) and resulting from the uncertainty
of $r_0/a$. In all other cases the error is statistical only.}
  \label{tab:all2}
\end{small}
\end{table}

\begin{table}[t!]
\begin{small}
   \centering
  \begin{tabular}[]{ccccccccccc}
    Ens. & $N$ & cnfs & step & $\langle A\rangle$ & $\tau_{int}$ & $\langle B\rangle$
& $\tau_{int}$ & $\langle C\rangle$ & $\tau_{int}$ & $r_0^4\chi$\\
\hline
A30.32 & 6 & 223 & 20 & 167.9(2.3) & 4.8(2.1) & 30.28(40) & 3.8(1.5) & -0.18(26) & 0.5(1) & 
0.0072(10)(3)(2)\\
A40.20 & 6 & 200 & 16 & 29.91(56) & 2.6(1.0) & 5.40(10) & 1.9(6) & -0.10(20) & 1.1(3) & 
0.0130(22)(5)(4)\\
A40.24 & 6 & 198 & 20 & 53.77(1.35) & 6.8(3.1) & 9.78(21) & 3.7(1.6) & -0.03(21) & 0.8(2) &
0.0086(13)(3)(2)\\
A40.32 & 6 & 190 & 16 & 170.5(1.2) & 1.9(6) & 30.74(20) & 1.3(4) & 0.25(34) & 0.7(2) &
0.0074(10)(3)(2)\\
A50.32 & 6 & 201 & 20 & 175.6(1.1) & 1.9(7) & 31.73(27) & 2.2(8) & 0.36(31) & 0.6(1) &
0.0081(12)(3)(2)\\
A60.24 & 6 & 163 & 8 & 54.46(59) & 1.6(5) & 9.87(13) & 1.4(5) & -0.26(25) & 0.9(3) & 0.0092(14)(3)(3)\\
A80.24 & 6 & 201 & 8 & 53.70(44) & 1.6(5) & 9.73(10) & 1.5(5) & 0.76(24) & 1.0(3) & 0.0114(17)(4)(3)\\
\hline
B25.32 & 8 & 199 & 20 & 91.88(1.27) & 1.7(6) & 18.95(23) & 1.3(4) & -0.57(32) & 1.2(3) &
0.0070(11)(1)(2)\\
B35.32 & 8 & 198 & 20 & 95.58(87) & 2.3(9) & 19.52(15) & 1.2(4) & -0.55(23) & 0.6(2) & 0.0067(9)(1)(2)\\
B55.32 & 6 & 538 & 40 & 95.59(34) & 1.1(2) & 19.47(7) & 0.9(2) & -0.06(16) & 0.6(1) & 0.0080(7)(2)(2)\\
B75.32 & 8 & 201 & 8 & 92.57(51) & 1.4(5) & 18.75(13) & 1.3(4) & 0.04(35) & 1.1(3) & 0.0090(11)(2)(3)\\
B85.24 & 12 & 236 & 20 & 31.48(20) & 0.8(2) & 6.52(5) & 0.6(1) & -0.09(15) & 0.7(1) & 0.0106(14)(2)(3)\\
\hline
D20.48 & 6 & 97 & 20 & 157.1(1.3) & 1.5(6) & 49.73(38) & 1.3(6) & -0.42(48) & 0.7(2) &
0.0041(11)(1)(1)\\
D30.48 & 6 & 101 & 20 & 158.2(9) & 1.0(4) & 50.18(24) & 0.6(2) & -0.44(64) & 0.9(3) & 0.0073(24)(1)(2)\\
D45.32 & 6 & 96 & 40 & 29.44(36) & 0.8(3) & 9.34(9) & 0.4(1) & -0.11(36) & 1.1(4) &
0.0125(21)(2)(4)\\
\hline
   \end{tabular}
  \caption{Our results for $N_f=2+1+1$ flavours. We give the ensemble label, the number of stochastic
sources $N$, the number of configurations used (cnfs), the step between measurements (in units of
molecular dynamics trajectories) and the values of
$\langle{\cal A}\rangle$, $\langle{\cal B}\rangle$, $\langle{\cal C}\rangle$ and the topological
susceptibility, together with integrated autocorrelation times $\tau_{int}$ (in units of
measurements).
The error for $r_0^4\chi$ is, respectively, statistical,
resulting from the uncertainty of $Z_P/Z_S$ (from the RI-MOM method) and resulting from the uncertainty
of $r_0/a$. In all other cases the error is statistical only.}
\label{tab:all211}
\end{small}
\end{table}

\begin{figure}[t!]
  \begin{center}
    \includegraphics[width=0.95\textwidth,angle=0]{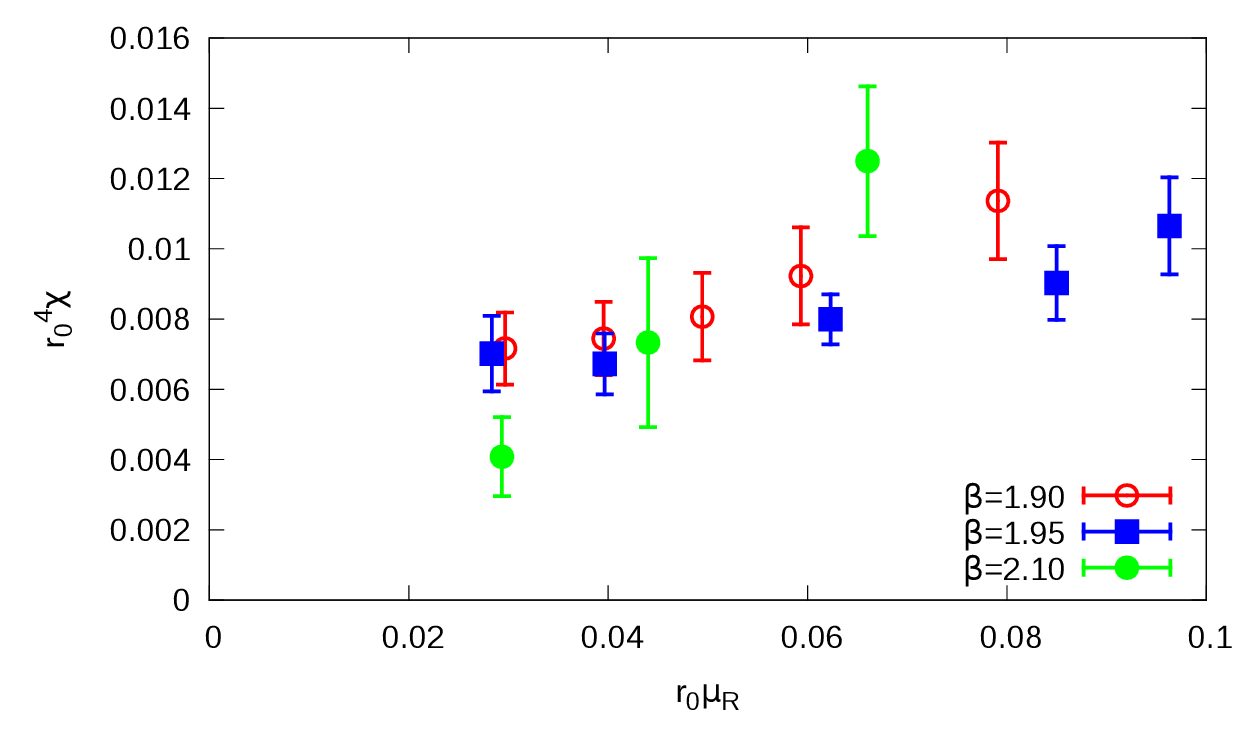}
  \end{center}
  \caption{\label{fig:all} The dependence of the dimensionless quantity $r_0^4\chi$ on the renormalized
quark mass $r_0\mu_R$. We show all ensembles used for the analysis of the $N_f=2+1+1$ flavour case.} 
\end{figure}

\subsection{$N_f=2+1+1$ results}
In this subsection, we discuss our data for the case with 2+1+1 active flavours. 
Our results for the observables $\langle{\cal A}\rangle$, $\langle{\cal
B}\rangle$, $\langle{\cal C}\rangle$ and the topological susceptibility are collected in
Tab.~\ref{tab:all211} and Fig.~\ref{fig:all} shows the results for the topological susceptibility.
We required that the autocorrelations for the topological charge are kept under control, i.e. can be
measured with reasonable accuracy using the method proposed in Ref.~\cite{Wolff:2003sm} (UW method).
This method allows for an estimate of the integrated autocorrelation time $\tau_{int}$ and of its
error.
We also made an independent error analysis using the method of bootstrap with blocking.
In all cases, we found results compatible with the UW method, given in Tab.~\ref{tab:all211}.
In particular, we found that the autocorrelation time for the observable ${\cal
C}$ is $\tau_{int}\lesssim1$.

Typically, we have $\mathcal{O}(200)$ configurations per ensemble, although for our ensembles at the
finest lattice spacing, we only have around 100 configurations.
Thus, the histograms that we can build have large statistical errors and within these large errors
the deviation from a zero-centered Gaussian is insignificant.
Few exceptions to this rule occur -- e.g. for ensemble A80.24 $\langle{\cal C}\rangle$ is more than
3$\sigma$ away from zero.
The typical error of the computed topological susceptibility is of the order of 15\% and we manage to go
below 10\% only for ensemble B55.32.
In this way, we conclude that the precision one can reach for the topological susceptibility is
only modest.
However, we want to emphasize here that this is a consequence of too short lengths of typical Monte
Carlo simulations in Lattice QCD and do not originate from the spectral projector method itself.
Especially with finer lattice spacings, autocorrelations are such that
to obtain truly independent gauge field configurations one has to perform measurements skipping
several trajectories.
Our experience shows that to obtain a 10\% precision in the computation of the topological
susceptibility,
we need around 300-400 truly independent configurations, which implies Monte Carlo runs of 10000-20000
trajectories, which is somewhat longer than is typically needed for most other applications.

\begin{figure}[t!]
\subfigure[$\beta=1.90$]{
\includegraphics[height=.29\textwidth]{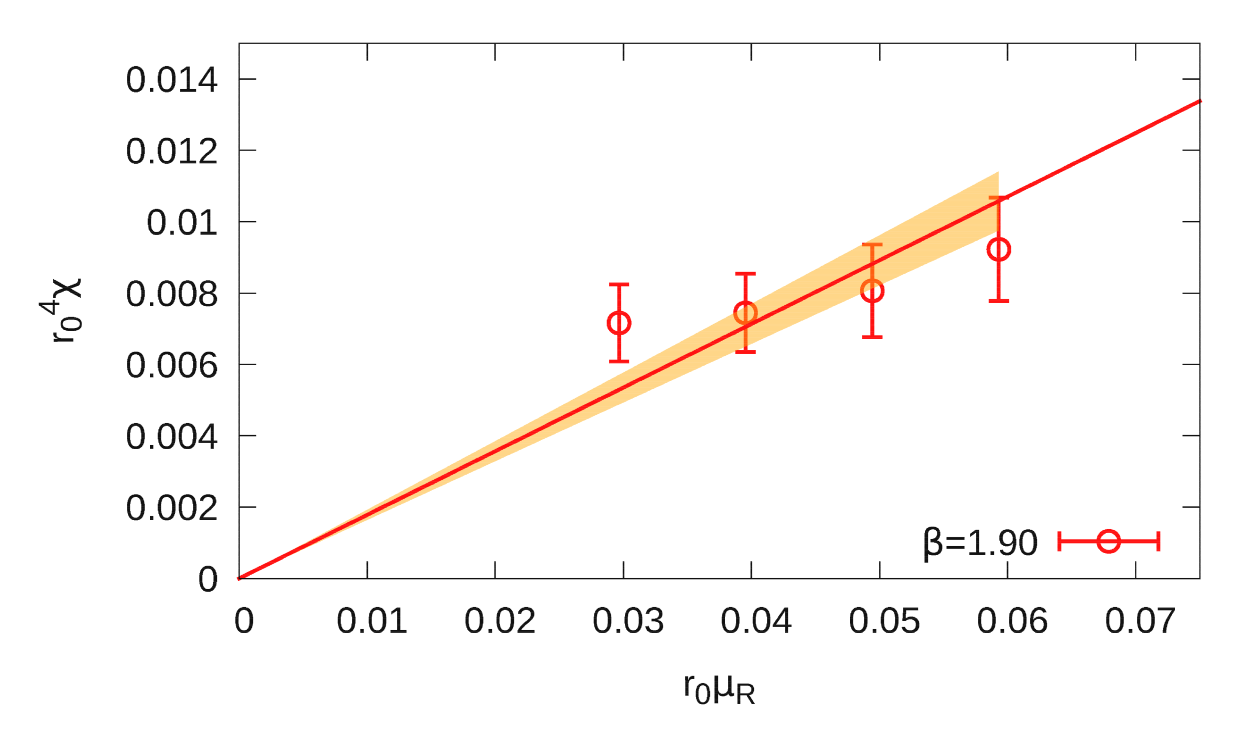}
}
\hspace*{-0.7cm}
\subfigure[$\beta=1.95$]{
\includegraphics[height=.29\textwidth]{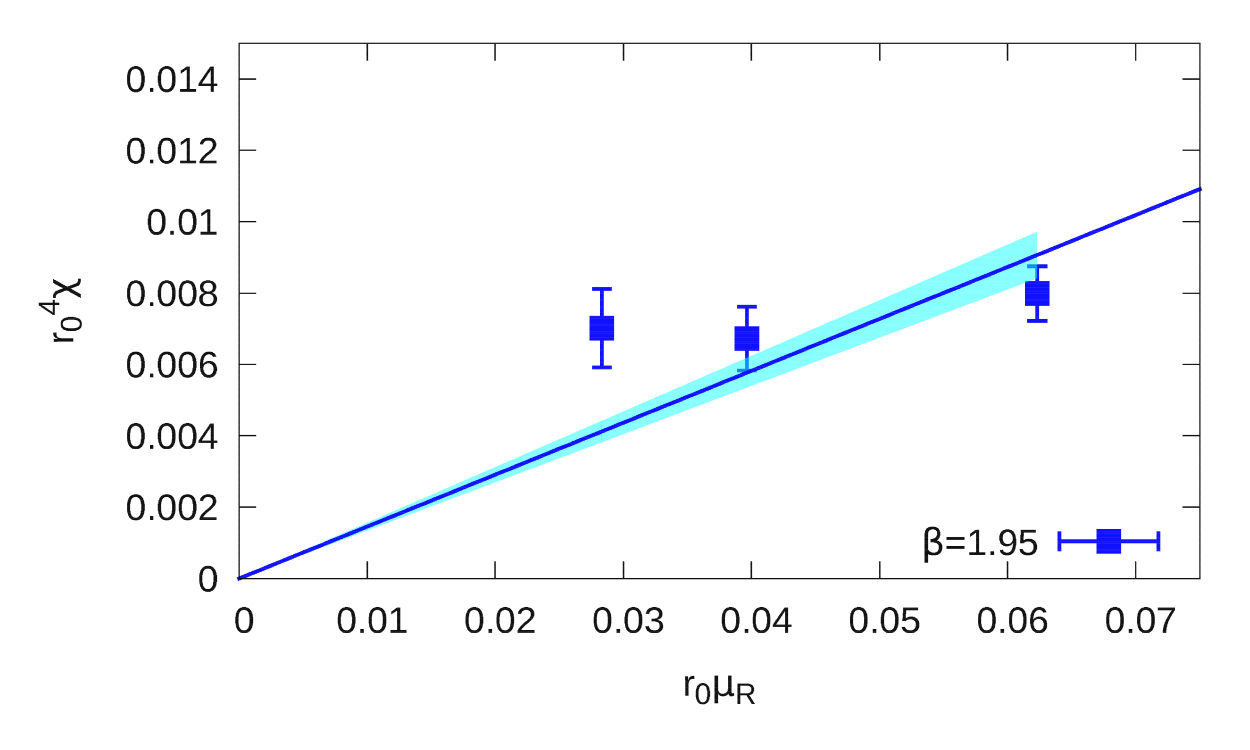}
}
\subfigure[$\beta=2.10$]{
\includegraphics[height=.29\textwidth]{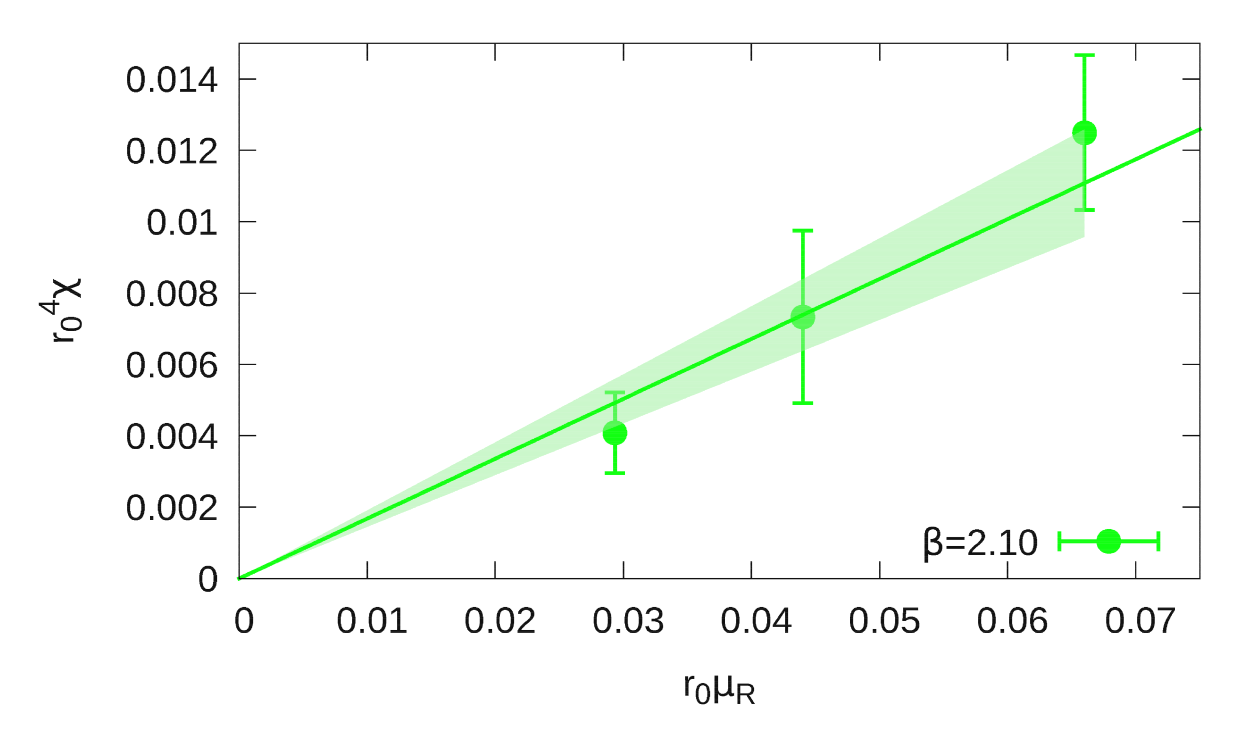}
}
\hspace*{-0.7cm}
\subfigure[continuum limit]{
\includegraphics[height=.29\textwidth]{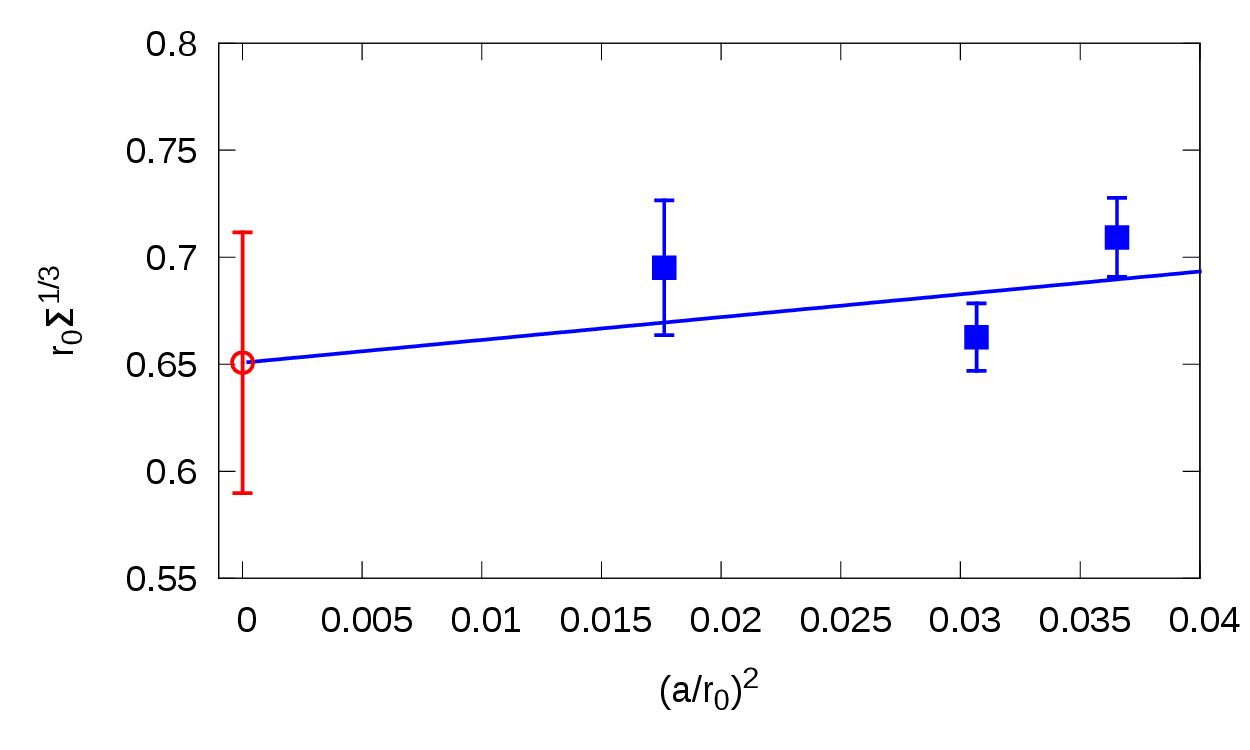}
}
\caption{(a,b,c) Renormalized quark mass dependence of the topological susceptibility for
$N_f=2+1+1$. The straight line corresponds to a fit of LO SU(2)
$\chi$PT. Only ensembles with $m_\pi\leq400$ MeV are included. $\chi^2/{\rm d.o.f.}$ values are:
1.45 (a), 5.02 (b), 0.49 (d). The continuum limit of
$r_0\Sigma^{1/3}$ extracted from fits shown in (a,b,c).}
\label{fig:fits}
\end{figure}

In order to describe the quark mass dependence of the topological susceptibility, we follow the same
strategy as discussed above for $N_f=2$ flavours.
Using only the LO$\chi$PT formula, we decided to apply a mass cut on our data,
excluding points for which the pion mass is larger than 400 MeV, i.e. keeping points for which
$r_0\mu_R<0.07$.
The fits of the LO formula to our data are shown in Fig.~\ref{fig:fits}(a,b,c).

As in the $N_f=2$ case, we can extract the chiral condensate from the dependence of $\chi$ on the quark
mass. 
We have performed an analysis separately for each lattice spacing, taking the individual
uncertainties of $Z_P/Z_S$, $Z_P$ and $r_0/a$ into account (in the way described in the previous section)
to propagate them to the values of $r_0\Sigma^{1/3}$ at finite lattice spacings (shown in
Fig.~\ref{fig:fits}(d)). Such obtained values are then extrapolated to the continuum limit, yielding the
value:
\begin{displaymath}
 r_0 \Sigma^{1/3} = 0.651(61).
\end{displaymath}
We mention here that it is possible to prove that the topological susceptibility computed using the
spectral projector method and twisted mass fermions at maximal twist is $\mathcal{O}(a)$-improved.
This is not guaranteed \emph{a priori} by standard arguments for the automatic
$\mathcal{O}(a)$-improvement at maximal twist \cite{Frezzotti:2003ni}, since the topological
susceptibility is defined via density chains that include integrals (in the continuum) or sums (on the
lattice) over all space time points, which leads to contact terms with short distance singularities. Such
contact terms can, in principle, spoil automatic $O(a)$-improvement.
The proof that this is not the case is sketched in Refs.~\cite{Cichy:2013lat,Cichy:2013egr}, while for
the details of this proof we refer to an upcoming publication \cite{letter}.

In general, the quality of our LO$\chi$PT fits is reasonable (see the values of $\chi^2/{\rm
d.o.f.}$ in the caption of Fig.~\ref{fig:fits}), with the exception of $\beta=1.95$, for which
$\chi^2/{\rm d.o.f.}\approx5$. This may signal the presence of effects beyond the ones captured in
our LO$\chi$PT fitting ansatz. However, with our current precision we are not able to address this
issue. The fact that some data points are off the fit line may well be a statistical fluctuation at this
level of precision.
As a check of the robustness of our result, we performed also another LO$\chi$PT fit including all our
data, i.e. also pion masses between 400 and 500 MeV. This leads to a
value for the chiral condensate in the continuum limit: $r_0 \Sigma^{1/3} = 0.619(58)$.
The result from this additional fit is slightly lower, although still compatible with
the one from fits applying a mass cut. The values of $\chi^2/{\rm d.o.f.}$ for the
LO$\chi$PT fits without pion mass cuts are: 1.70 ($\beta=1.9$), 4.52 ($\beta=1.95$), 0.49 ($\beta=2.1$),
i.e. they are comparable to the ones for fits without pion mass cuts (see the caption of
Fig.~\ref{fig:fits}).

The error that we give is dominated by statistical uncertainties, but the contribution from the
systematic errors related to $r_0/a$ and $Z_P/Z_S$ is also included.
However, it does not include the main source of systematic effects coming from $\chi$PT: the use of
only the leading order expression.
As we mentioned above, our precision is not enough to use an NLO fitting ansatz.
Still, our result is in agreement with the direct determination from the mode number on the same set of
gauge field ensembles -- $r_0 \Sigma^{1/3} = 0.680(20)(21)$ \cite{Cichy:2013gja}, indicating that
LO$\chi$PT describes the quark mass dependence of the topological susceptibility at least within the
rather large errors of our results.

It is worth emphasizing that at $\beta=1.9$ and $\beta=1.95$ the data for $r_0^4\chi$ do not show a
clear tendency to assume a zero value when the quark mass is decreased. Only at $\beta=2.1$ and hence
closer to the continuum limit, the data seem to approach zero linearly in the quark mass. Thus, in order
to cleanly identify this expected behaviour of the topological susceptibility, smaller quark masses and
a significantly increased precision are required.

\section{Conclusions}
We have computed the topological susceptibility in dynamical Lattice QCD
simulations using the method of spectral projectors.
This method has two important advantages that we want to emphasize here:
\begin{itemize}
 \item it relies on a theoretically sound definition of the topological susceptibility from density
chain correlators that is free of short distance singularities,
\item it is significantly less computer time expensive than the topological susceptibility
computation from the index of the overlap Dirac operator.
\end{itemize}
One main result of our work is that the topological susceptibility is affected by substantial
statistical fluctuations necessitating long Monte Carlo histories. 
With typical parameter values of Lattice QCD simulations nowadays, i.e. lattice spacings of 0.05$\,$fm
$\lesssim$ $a$ $\lesssim$ $0.1\,$fm and lengths of Monte Carlo runs of $\mathcal{O}(5000)$ trajectories
with autocorrelation times \mbox{$\tau_{int}=\mathcal{O}(10)$} trajectories, it is very difficult to
obtain errors smaller than 10-15\% for a given ensemble.
We emphasize that this is not a property of the method used here, but of the gauge field
configurations themselves and as such can not be easily overcome, i.e. without running very long
simulations. In addition, the topological
properties of gauge fields -- here characterized by the quantity ${\cal C}$ of Eq.~\eqref{eq:C}, which
is closely related to the topological charge -- tend to be particularly susceptible to autocorrelation
effects, which increase with decreasing lattice spacing.
This is indeed observed with the present method and implies that very high statistics is needed (in
particular at small lattice spacings) to
overcome this problem, unless one works with open boundary conditions that naturally allow to move the
problem to at present unachievably small lattice spacings \cite{Schaefer:2010aa}.

Despite these difficulties, we were able to demonstrate that by \emph{imposing} LO chiral perturbation
theory as a description of our data for the topological susceptibility, values of the chiral condensate
could be determined, which read: $r_0\Sigma^{1/3}=0.650(22)$ ($N_f=2$, no continuum extrapolation) and
$r_0 \Sigma^{1/3} = 0.651(61)$ ($N_f=2+1+1$).
These results, although having large errors for the reasons discussed above, are fully compatible with
the ones of our direct calculation using spectral projectors \cite{Cichy:2013gja}.
We estimate that a meaningful test of the NLO chiral perturbation theory prediction for the quark
mass dependence of the topological susceptibility would require a factor 3-10 longer runs (than typical
ones, as specified above), which would bring the errors down below 10\%.
Nevertheless, we have shown that such calculations are becoming feasible with present-day computing
resources and the advantages of computations with a theoretically sound definition of the topological
susceptibility using density chains, promote the here used method to one of the most promising ways to
address topological properties of QCD in the future.

\vspace{0.3cm}
\noindent {\bf Acknowledgments} We thank the European
Twisted Mass Collaboration for generating gauge field ensembles used in this work.
We are grateful to A.~Shindler for collaboration and discussions concerning $\mathcal{O}(a)$ improvement
of the topological susceptibility.
We acknowledge useful discussions with V.~Drach, G.~Herdoiza, M.~M\"uller-Preussker, K.~Ottnad,
G.C.~Rossi, C.~Urbach, F.~Zimmermann.
K.C. was supported by Foundation for Polish Science fellowship ``Kolumb''.
This work was supported in part by the DFG Sonderforschungsbereich/Transregio SFB/TR9. 
K.J. was supported in part by the Cyprus Research Promotion
Foundation under contract $\Pi$PO$\Sigma$E$\Lambda$KY$\Sigma$H/EM$\Pi$EIPO$\Sigma$/0311/16.
The computer time for this project was made available to us by the J\"ulich
Supercomputing Center, LRZ in Munich, the PC cluster in Zeuthen, Poznan Supercomputing and Networking
Center (PCSS). We thank these computer centers and their staff for all technical advice and help.

\appendix
\section{Number of stochastic sources}
\label{sec:Nsrc}
\begin{figure}[t!]
  \begin{center}
\includegraphics[width=0.345\textwidth,angle=270]{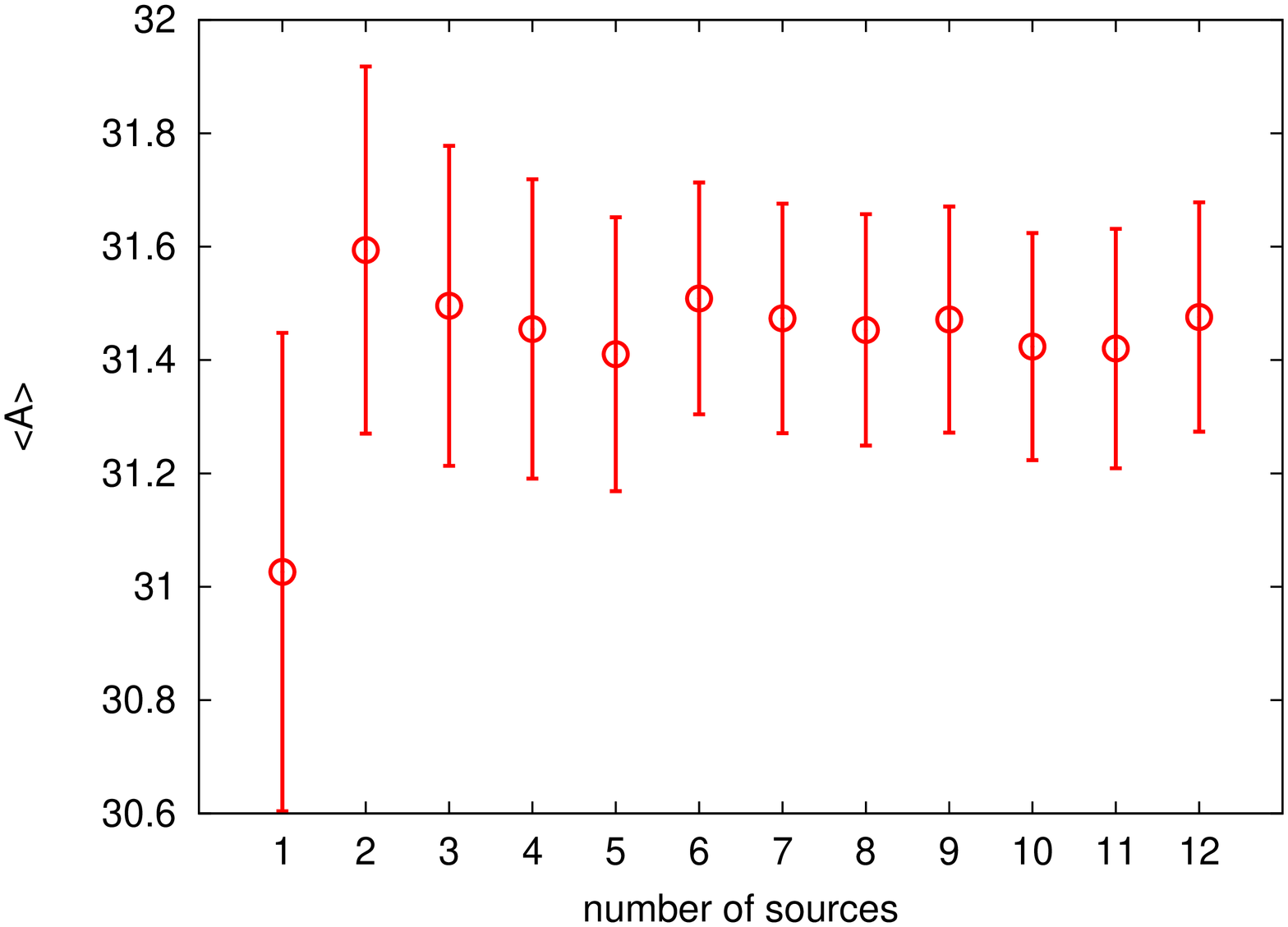}
\includegraphics[width=0.345\textwidth,angle=270]{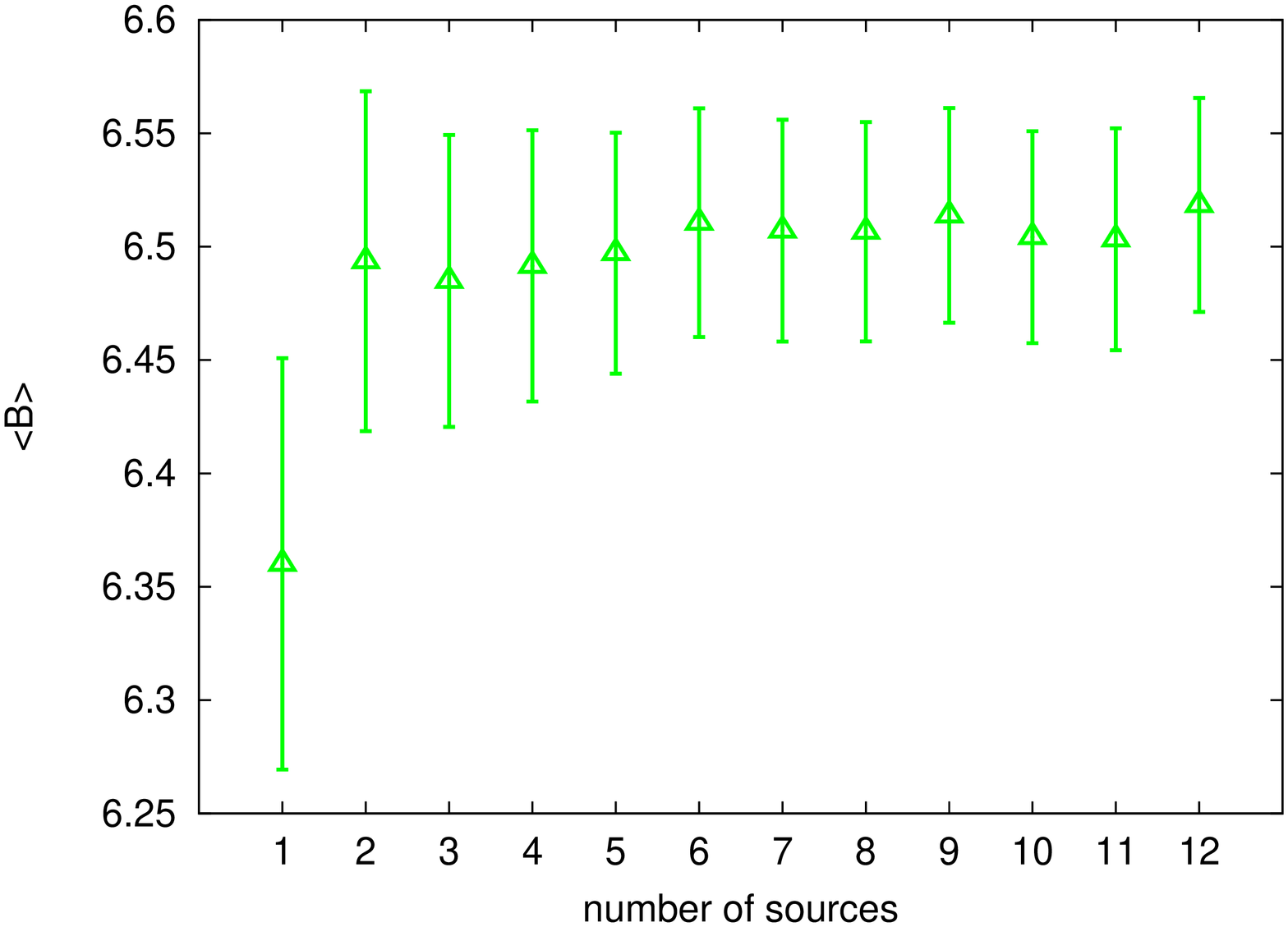}
\includegraphics[width=0.345\textwidth,angle=270]{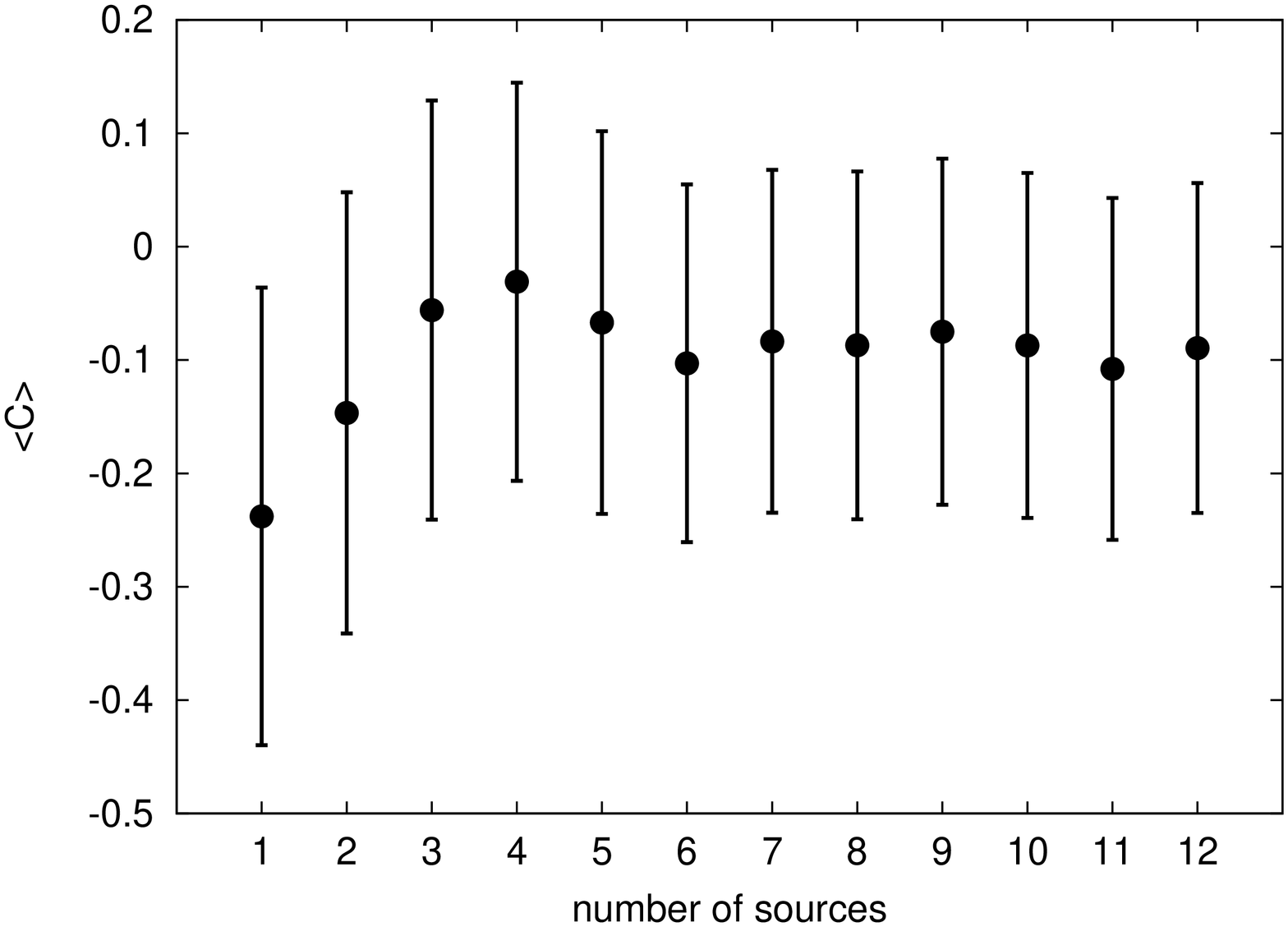}
\includegraphics[width=0.345\textwidth,angle=270]{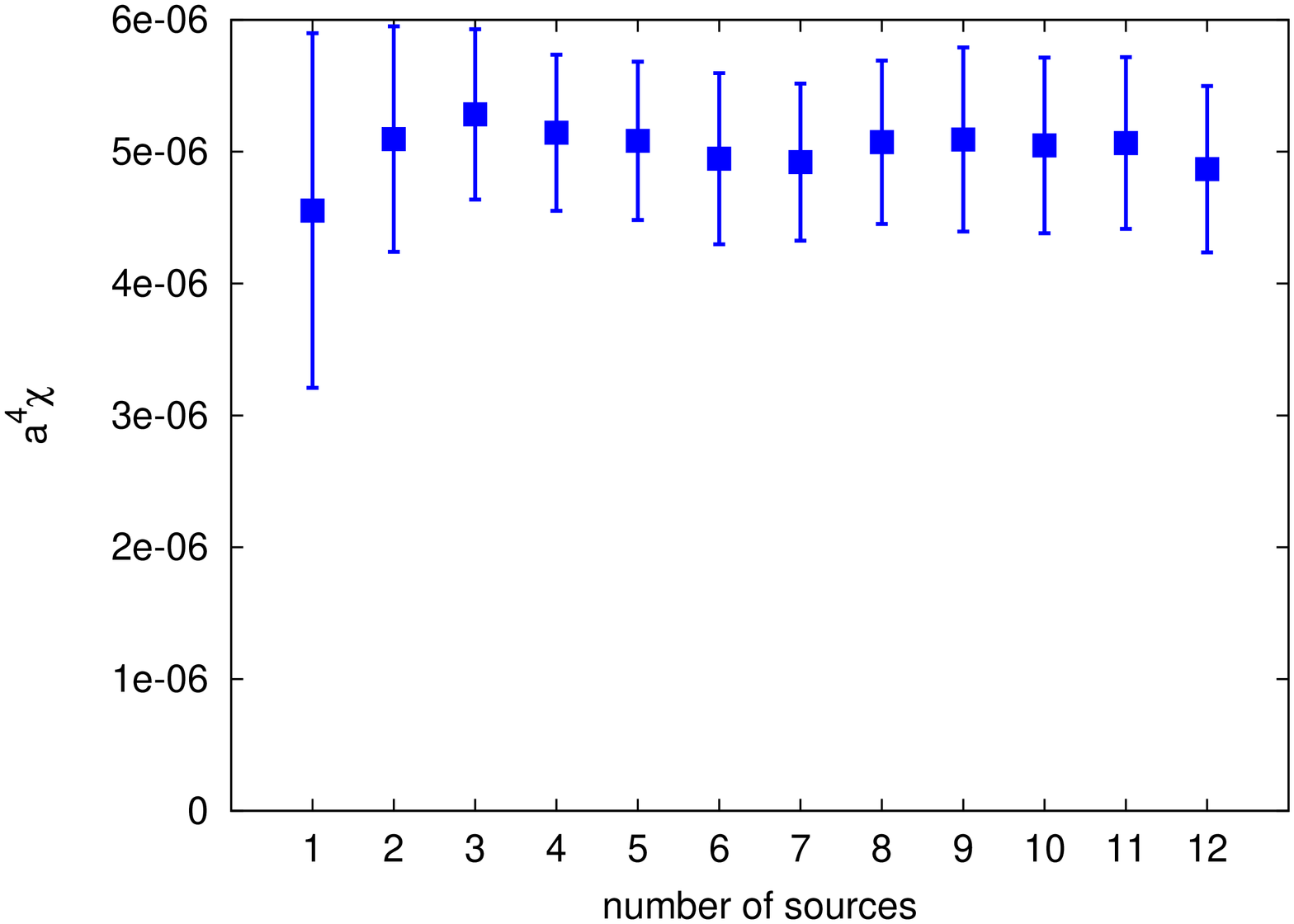}
  \end{center}
  \caption{\label{fig:Nsrc} Dependence of the computed observables $\langle{\cal
A}\rangle$, $\langle{\cal B}\rangle$, $\langle{\cal C}\rangle$ and the bare topological susceptibility
$a^4\chi$ on the number of stochastic sources. Ensemble B85.24.} 
\end{figure}

The bare topological susceptibility is given by the formula $(\langle{\cal C}^2\rangle-\langle{\cal
B}\rangle)/N$, where the number of stochastic sources $N$ enters explicitly.
To achieve good precision for the topological susceptibility, it is desirable to have as many stochastic
sources as possible.
This, however, of course increases the numerical cost of the computation.
According to Ref.~\cite{Luscher:2010ik}, a sensible compromise is achieved if $N=6$.
We have investigated this issue numerically for ensemble B85.24 and our results are presented in
Fig.~\ref{fig:Nsrc}.
Apart from the observable $\langle{\cal C}\rangle$, which has an error basically independent on $N$
(with a slight reduction of the error by adding a second source), the other observables show a similar
pattern -- the error reduces considerably by adding a second and third source and then it still
decreases, but more slowly.
Finally, our conclusion agrees with the one of Ref.~\cite{Luscher:2010ik} that $N=6$ is a reasonable
compromise.
To reduce the statistical error if one already has 6 stochastic sources per configuration, it is
more advisable to
increase the number of independent gauge field configurations.
Therefore, we decided to use $N=6$ for most of our computations -- only in some cases when it was not
possible to increase statistics by adding more configurations, we decided to increase the number of
stochastic sources to 8 or 12.

\bibliographystyle{jhep}
\bibliography{top_susc}	

\end{document}